\documentclass[letterpaper,twocolumn,10pt]{article}
\usepackage{amsmath,amssymb}
\usepackage{usenix-2020-09}

\usepackage{booktabs}
\usepackage{longtable}
\usepackage{xltabular}
\usepackage{tabularx}
\usepackage{array}
\usepackage{multirow}
\usepackage{colortbl}
\usepackage{float}
\usepackage{graphicx}
\usepackage{tikz}
\usetikzlibrary{arrows.meta,positioning,fit,patterns}
\usepackage{pgfplots}
\pgfplotsset{compat=1.18}
\usepackage{enumitem}
\usepackage{xurl}   %
\usepackage[labelfont=bf,skip=4pt]{caption}  %
\makeatletter
\renewcommand\paragraph{\@startsection{paragraph}{4}{\z@}%
  {2ex \@plus 0.8ex \@minus .3ex}{-1em}{\normalfont\normalsize\bfseries}}
\makeatother

\usepackage{listings}
\newfloat{algorithm}{tp}{loa}
\floatname{algorithm}{Algorithm}
\lstdefinestyle{mesh}{basicstyle=\footnotesize, columns=fullflexible, literate={-}{{-}}1,
  keepspaces=true, mathescape=true, numbers=left, numberstyle=\tiny,
  numbersep=4pt, xleftmargin=10pt, aboveskip=2pt, belowskip=0pt,
  morekeywords={for,if,else,while,return,each,in,per,parallel},
  keywordstyle=\bfseries, morecomment=[l]{//}, commentstyle=\itshape,
  frame=none, showstringspaces=false}

\newif\ifreview
\ifdefined\REVIEWBUILD\reviewtrue\else\reviewfalse\fi
\definecolor{ProposalBlue}{RGB}{0,67,148}
\definecolor{CorrectionPurple}{RGB}{118,45,145}
\definecolor{DecisionOrange}{RGB}{155,72,0}
\definecolor{EvidenceShade}{RGB}{239,243,248}

\newenvironment{corrected}{\ifreview\color{CorrectionPurple}\fi}{}
\newcommand{\correction}[1]{{\ifreview\color{CorrectionPurple}\fi#1}}
\newif\ifclarity
\ifdefined\CLARITYBUILD\claritytrue\else\clarityfalse\fi
\ifclarity\usepackage[normalem]{ulem}\fi
\definecolor{ClarityGreen}{RGB}{0,90,25}
\DeclareRobustCommand{\clarity}[1]{#1}
\newenvironment{clarityblock}{}{}
\newif\ifanon
\anonfalse

\newcommand{\EvalBoardCorridors}{37}

\newcommand{\EvalBoardTierMatches}{501}

\newcommand{\EvalBoardDecisionMatches}{507}

\newcommand{\EvalBoardAnalyzerDecided}{536}
\newcommand{\EvalBoardAgentAgreementPct}{94.6}

\newcommand{\CorpusN}{72}
\newcommand{\CorpusAdjacentN}{5}
\newcommand{\CorpusCatalogueN}{77}

\newcommand{\CorpusPostureN}{29}

\newcommand{\CorpusGrossB}{3.52}

\newcommand{\CorpusAdjustedPct}{75}
\newcommand{\CorpusRecentPct}{88}

\newcommand{\CorpusRecentPostureN}{20}

\newcommand{\CorpusWithoutKelpPct}{82}
\newcommand{\CorpusBoundaryPct}{87}

\newcommand{\CorpusPostureM}{1{,}588.5}
\newcommand{\CorpusCapsN}{28}

\newcommand{\CorpusBasisReadN}{11}

\newcommand{\CorpusTimingN}{7}
\newcommand{\CorpusTimingWithinHourN}{4}

\newcommand{\CorpusPreExploitCapsN}{21}
\newcommand{\CorpusPreExploitCapsM}{1{,}339.2}

\newcommand{\CorpusAuthorityN}{11}

\newcommand{\CorpusQuorumN}{17}

\newcommand{\CorpusBackingN}{7}

\newcommand{\CorpusProvenanceN}{6}

\newcommand{\CorpusClassQuorumN}{11}
\newcommand{\CorpusClassQuorumM}{1{,}095}
\newcommand{\CorpusClassAdminN}{10}
\newcommand{\CorpusClassAdminM}{176}
\newcommand{\CorpusClassAttesterN}{7}
\newcommand{\CorpusClassAttesterM}{318}
\newcommand{\CorpusNearMissN}{7}

\newcommand{\CorpusClassOtherM}{0.2}
\newcommand{\IdealControlResidualM}{127}

\newcommand{\IdealControlReductionFold}{27}
\newcommand{\IdealControlPreventedN}{25}
\newcommand{\IdealControlBoundedN}{30}

\newcommand{\IdealControlZeroN}{8}
\newcommand{\IdealControlBoundedM}{31.251}

\newcommand{\IdealControlAllowanceM}{1}
\newcommand{\IdealControlRefillM}{1}

\newcommand{\IdealControlStopHours}{1}

\newcommand{\IdealControlDayReductionFold}{22}
\newcommand{\PhysicalTierBranches}{48}
\newcommand{\EvidenceTierBoundaries}{34}
\newcommand{\UnestablishedTierBoundaries}{2}

\newcommand{\HistoricalFilteredAgentN}{8}

\providecommand{\IdealControlResidualM}{}\renewcommand{\IdealControlResidualM}{33}

\providecommand{\IdealControlReductionFold}{}\renewcommand{\IdealControlReductionFold}{105}
\providecommand{\IdealControlPreventedN}{}\renewcommand{\IdealControlPreventedN}{28}
\providecommand{\IdealControlBoundedN}{}\renewcommand{\IdealControlBoundedN}{36}
\providecommand{\IdealControlBoundedM}{}\renewcommand{\IdealControlBoundedM}{33.0}

\providecommand{\IdealControlOutsideM}{}\renewcommand{\IdealControlOutsideM}{0.2}

\providecommand{\PostureFailuresN}{}\renewcommand{\PostureFailuresN}{29}
\providecommand{\LossHacksN}{}\renewcommand{\LossHacksN}{65}
\providecommand{\LossCodeN}{}\renewcommand{\LossCodeN}{36}

\providecommand{\IdealControlZeroN}{}\renewcommand{\IdealControlZeroN}{6}

\providecommand{\IdealControlDayReductionFold}{}\renewcommand{\IdealControlDayReductionFold}{59}
\providecommand{\IdealControlMinFold}{}\renewcommand{\IdealControlMinFold}{18}

\providecommand{\ContainmentCreditResidualM}{}\renewcommand{\ContainmentCreditResidualM}{120}
\providecommand{\ContainmentCreditFold}{}\renewcommand{\ContainmentCreditFold}{29.3}

\providecommand{\ContainmentCreditPreventedN}{}\renewcommand{\ContainmentCreditPreventedN}{6}

\providecommand{\RateStopResidualM}{}\renewcommand{\RateStopResidualM}{123}
\providecommand{\RateStopFold}{}\renewcommand{\RateStopFold}{28.6}

\providecommand{\CoverageNamedStopsN}{}\renewcommand{\CoverageNamedStopsN}{41}
\providecommand{\CoverageNamedMissN}{}\renewcommand{\CoverageNamedMissN}{3}

\providecommand{\CoverageUnnamedContainedN}{}\renewcommand{\CoverageUnnamedContainedN}{26}

\providecommand{\CoverageTotalN}{}\renewcommand{\CoverageTotalN}{72}
\providecommand{\CapCheckCappedN}{}\renewcommand{\CapCheckCappedN}{28}

\newcommand{\artifact}[1]{}

\newcommand{\tier}[2]{\textbf{#1}\,#2\hspace{7pt plus 3pt minus 2pt}}
\newcommand{\tiers}[1]{#1\unskip}

\title{Lessons Not Learned:\\
Why Cross-Chain Bridges Keep Losing Billions the Same Way}

\newif\ifshowauthors
\showauthorstrue
\ifshowauthors
\author{
{\rm Mohsen Ahmadvand, Piotr Dobaczewski and Camille Bouzerand}\\
Monad Foundation
}
\else
\author{{\rm Anonymous submission}}
\fi

\begin{document}
\maketitle

\begin{abstract}
\clarity{Bridge hacks have cost billions, often without exploiting a code bug. In the
cross-chain bridge hacks we catalogue from 2021 to 2026, failures of security
posture, meaning how a bridge's people, code, controls and off-chain
infrastructure are set up, account for \CorpusRecentPct{}\% of reported losses since 2023,
and nearly all repeat earlier failures.

Code audits cannot catch these failures, so we propose a bridge security
assessment rubric. It
scores each bridge deployment from on-chain evidence, and a hard cap drops the
score to the lowest value when a fatal risk is present. A token is only as
safe as its weakest deployment if all its deployments share one backing.
For attacks nobody anticipated, we show that rate limits and an emergency stop
still bound the loss. We build an analyzer that applies the rubric to deployed
bridges and monitors them for changes, and an agentic analyzer that runs the
same checks on any bridge. Operators learn what to fix, and users can compare
bridges.

Our rubric's controls cut bridge hack losses
\IdealControlReductionFold{}-fold, from \$\CorpusGrossB{}B to
\$\IdealControlResidualM{}M.}
\end{abstract}

\section{Introduction}
\label{sec:intro}
\begingroup
The April 2026 KelpDAO incident shows that a bridge with audited code can
still lose its funds through its configuration. KelpDAO's bridge used LayerZero's Omnichain Fungible Token (OFT)
contracts, which had been audited repeatedly. LayerZero's report attributes
the \$292M loss to one deployment that accepted messages approved by a single
decentralized verifier network (DVN)~\cite{pm-kelpdao}. The attacker defeated
that verifier by poisoning its remote procedure call (RPC) feed. The single-DVN
setting was public on-chain before the exploit, so anyone could have seen the weakness
before committing funds (Section~\ref{sec:preexploit}).

Configuration matters because a cross-chain bridge authorizes a release or
mint on one blockchain using evidence from
another~\cite{zamyatin2021sok,belchior2021survey,mccorry2021sok}. Its security
depends on the contracts, keys and services that verify that
evidence~\cite{zhang2024sok}. A code audit cannot show whether those keys and
services are safely set up. Attackers drained Ronin with stolen
keys~\cite{pm-ronin}; at Multichain, one party controlled the whole signing
system and could move the funds alone~\cite{pm-multichain}. We call these failures of \emph{security
posture}, meaning how a bridge's people, code, controls and off-chain
infrastructure are set up.

Before 2023, most reported losses came from code defects; since 2023, posture failures account for \CorpusRecentPct{}\% of reported losses (Figure~\ref{fig:trend}).
Assessment must therefore establish who can authorize a release, what
evidence they need, and how much can leave if they are compromised.

We catalogue \CorpusN{} bridge incidents with \$\CorpusGrossB{}B in reported
gross losses~\cite{rekt,defillama,chainalysis2022} and build from their
recurring failures a bridge security assessment rubric. It scores each
deployment from on-chain evidence, and a hard cap drops the score to the
lowest value when a fatal risk is present. An asset scores as its weakest
component on any chain that can mint it or release its backing, because when
deployments share one backing, one such component can empty it for every
chain (Section~\ref{sec:rubric}). We implement the checks in an analyzer and
extend it to live monitoring, so assessments stay current as bridges change.

A future attack may exploit a weakness nobody
anticipated, so we also show that a rate limit and an emergency stop bound
the loss of a successful attack, as long as the attacker cannot remove them
(Section~\ref{sec:damage}); we call these controls \emph{containment}. Code bugs are outside our rubric,
as are economic attacks and issuer powers, but rate limits and an emergency
stop cap losses from any attack, so our evaluation applies them to code bugs
too.

L2BEAT, DeFiSafety and the Crosschain Risk Framework rate bridges by design
trust, process quality or worst-category risk~\cite{l2beat,defisafety,crosschainrisk}, and
contract-assurance standards address implementation
quality~\cite{eea_ethtrust,owasp_scsvs}. Our rubric differs in two ways: every
critical check must be decided by evidence, and an asset is scored across its
deployments on every chain (Section~\ref{sec:related}).
\endgroup

\definecolor{BBposture}{HTML}{2A78D6}
\definecolor{BBcode}{HTML}{EB6834}
\definecolor{BBrule}{HTML}{9BA3AE}
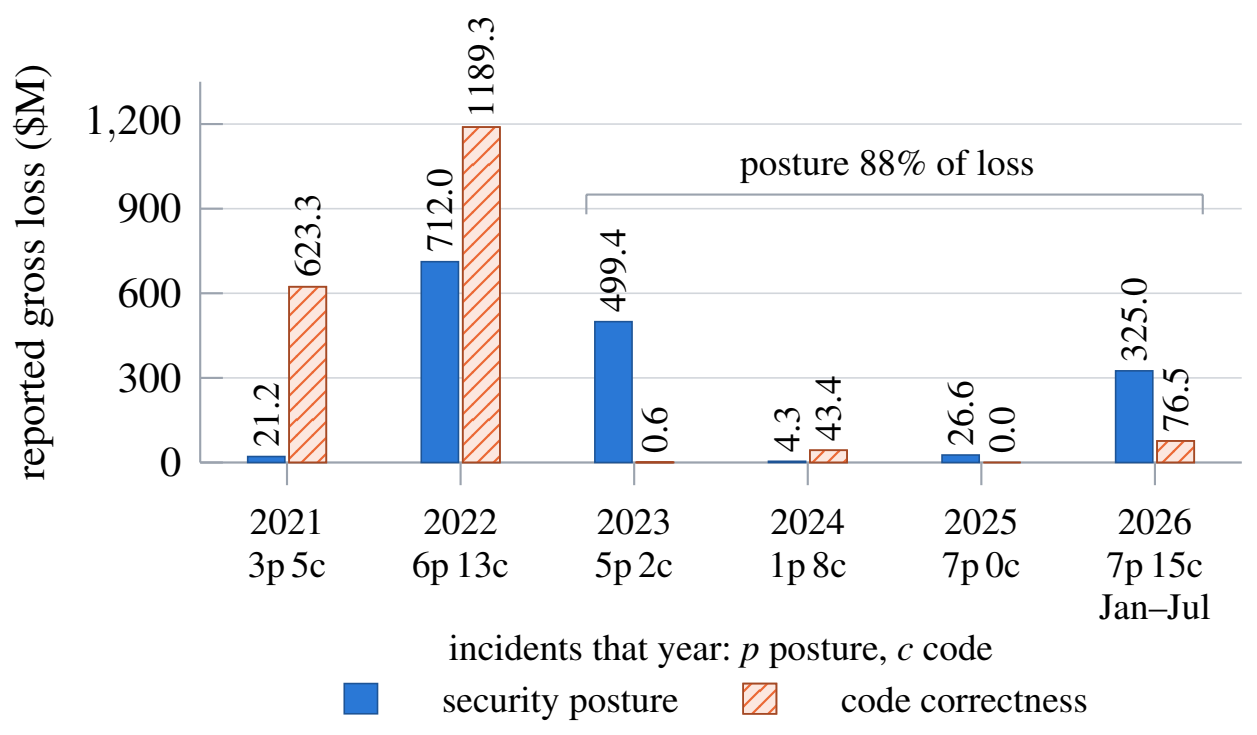
\begin{figure}[tp]
\centering
\begin{tikzpicture}
\begin{axis}[
  width=\columnwidth, height=4.1cm,
  ybar=0.8pt, bar width=7pt,
  ymin=0, ymax=1350,
  ytick={0,300,600,900,1200},
  yticklabel style={font=\footnotesize},
  ylabel={reported gross loss (\$M)},
  ylabel style={font=\footnotesize, yshift=-1pt},
  xlabel={incidents that year: \textit{p} posture, \textit{c} code},
  xlabel style={font=\scriptsize, yshift=4pt},
  symbolic x coords={2021,2022,2023,2024,2025,2026},
  xtick=data,
  xticklabels={{2021\\3p\,5c},{2022\\6p\,13c},{2023\\5p\,2c},{2024\\1p\,8c},{2025\\7p\,0c},{2026\\7p\,15c\\Jan--Jul}},
  xticklabel style={font=\scriptsize, align=center, yshift=-1pt},
  enlarge x limits=0.10,
  axis lines*=left,
  axis line style={draw=BBrule, line width=0.4pt},
  tick style={draw=BBrule, line width=0.4pt},
  ymajorgrids, grid style={draw=BBrule!45, line width=0.3pt},
  clip=false,
  nodes near coords, point meta=explicit symbolic,
  every node near coord/.append style={font=\scriptsize, rotate=90,
    anchor=west, inner sep=1.6pt},
  legend style={font=\scriptsize, draw=none, fill=none, at={(0.5,-0.53)},
    anchor=north, legend columns=2, column sep=10pt},
  legend image code/.code={\draw[#1] (0cm,-0.07cm) rectangle (0.22cm,0.13cm);},
]
\addplot[draw=BBposture!70!black, line width=0.3pt, fill=BBposture]
  coordinates {(2021,21.2) [21.2] (2022,712.0) [712.0] (2023,499.4) [499.4] (2024,4.3) [4.3] (2025,26.6) [26.6] (2026,325.0) [325.0]};
\addlegendentry{security posture}
\addplot[draw=BBcode!70!black, line width=0.3pt, fill=BBcode!16,
         postaction={pattern=north east lines, pattern color=BBcode}]
  coordinates {(2021,623.3) [623.3] (2022,1189.3) [1189.3] (2023,0.6) [0.6] (2024,43.4) [43.4] (2025,0.0) [0.0] (2026,76.5) [76.5]};
\addlegendentry{code correctness}
\draw[BBrule, line width=0.4pt]
  ([xshift=-9pt]axis cs:2023,950) -- ([xshift=9pt]axis cs:2026,950)
  ([xshift=-9pt]axis cs:2023,950) -- ([xshift=-9pt]axis cs:2023,880)
  ([xshift=9pt]axis cs:2026,950) -- ([xshift=9pt]axis cs:2026,880);
\node[font=\scriptsize, anchor=south, inner sep=2pt, xshift=15pt]
  at (axis cs:2024,950) {posture \CorpusRecentPct{}\% of loss};
\end{axis}
\end{tikzpicture}
\caption{Posture failures dominate reported bridge losses since 2023; early losses concentrate in code defects. Bars show reported gross loss per year, with incident counts below the axis. The 2026 window covers January--July. \clarity{Section~\ref{sec:rq1}} tests sensitivity to the largest recent loss and classification choices.}
\label{fig:trend}
\end{figure}

\paragraph{Findings.}
\begingroup
Bridges keep failing the same way: every posture failure since 2023
repeats a kind seen by 2022 (Section~\ref{sec:rq1}). Our rubric flags all
\CorpusPostureN{} posture failures: a hard cap fires on \CorpusCapsN{}, and the
last, Celer's hijacked website, gets the lowest tier of our website question
(Section~\ref{sec:caps}). For \CorpusPreExploitCapsN{} of the
\CorpusPostureN{}, the weakness was visible on chain before the attack, and the
other \the\numexpr\CorpusPostureN-\CorpusPreExploitCapsN\relax{} were off-chain
(Section~\ref{sec:preexploit}). Our scores have real impact: higher tiers
lower losses or make bridges verifiable (Section~\ref{sec:validation}). Our
rubric's controls cut bridge hack losses \IdealControlReductionFold{}-fold
(Section~\ref{sec:rq4}).
\endgroup

\paragraph{\clarity{Implications.}}
\clarity{Operators should keep the keys that move assets apart from the keys
that stop the bridge, limit every path that can release funds, and monitor
changes to these keys and limits. Users, applications and institutions can
compare bridges by who can move their funds and which controls limit a loss.}

\paragraph{Contributions.}
\begin{itemize}[noitemsep,topsep=2pt,leftmargin=1.4em]
\item \clarity{We catalogue \CorpusN{} bridge incidents from 2021 to 2026 and trace
  each loss to the component that failed (Section~\ref{sec:catalogue}).}
\item \clarity{We build a counterfactual model that replays each hack under
  different defenses (Section~\ref{sec:eval-model}). With it, we show that rate
  limits and an emergency stop cap what an attacker can take; together with
  our rubric's posture controls, they cut bridge hack losses
  \IdealControlReductionFold{}-fold (Sections~\ref{sec:damage}
  and~\ref{sec:rq4}).}
\item \clarity{Our rubric introduces a metric to quantify the security of bridges
  through tiers (scores) and hard caps that indicate fatal risks. With the same
  model, we show that the tiers meaningfully control losses
  (Sections~\ref{sec:rubric} and~\ref{sec:validation}).}
\item \clarity{Our approach assesses risk across chains and tracks it at the
  weakest leg, coupled with live monitoring; we evaluate 37 tokens listed on Monad (Sections~\ref{sec:system} and~\ref{sec:rq5}).}
\item \clarity{We build an agentic analyzer that runs the rubric's checks on
  any bridge; it gives the analyzer's answer on \EvalBoardAgentAgreementPct{}\%
  of the answers the analyzer decides and re-checks an unchanged bridge at no further
  model cost
  (Sections~\ref{sec:agentic} and~\ref{sec:rq7}).}
\end{itemize}

\clarity{Section~\ref{sec:background} introduces the terms the rest of the
paper uses and the gaps in prior work we fill. Section~\ref{sec:design} presents our design: it catalogues the
incidents and the failures behind them, shows how to limit losses when
prevention fails, and turns these lessons into our bridge security
assessment rubric. Section~\ref{sec:system} implements the rubric in our
analyzer, extended to live monitoring, and in our agentic analyzer, which runs
the same checks on any bridge. Section~\ref{sec:eval} tests each
claim against the catalogue and live bridges. Section~\ref{sec:discussion}
draws lessons for operators, and Section~\ref{sec:limits} states the limitations
and threats to validity.}

\section{Background and Related Work}
\label{sec:background}
\begin{clarityblock}
This section describes what a bridge relies on, the two tools we use to model
bridge security and the attacker we assume. It then reviews prior work and the
gap each part of this paper fills.

\subsection{Bridges}
A bridge moves a token between chains by locking it in escrow and minting a
copy~\cite{wormhole,polygonpos}, by burning and re-minting it~\cite{cctp}, or
by paying out from funds already on the
destination~\cite{hop,cbridge,stargate,synapse,across,debridge}. The
destination chain cannot read the source chain, so it must trust someone to
report what happened there~\cite{zamyatin2021sok}.

\paragraph{Dependencies.}
A bridge relies on four things: people, code, controls and off-chain
infrastructure. \emph{People} hold the bridge's keys: the upgrade key, signer
keys and operator keys. An externally owned account (\emph{EOA}) key acts
alone, a \emph{multisig} needs several keys, and a \emph{timelock} delays an
action so defenders can respond. \emph{Code} enforces the rules: how many
signatures a release needs, who may mint, what backs the token, which
contracts on other chains to accept messages from, and that no message is used
twice. Whoever holds a proxy's upgrade key can replace the code.
\emph{Controls} limit the damage once something fails: rate limits, limits on
total supply, an emergency stop and monitoring of changes. \emph{Off-chain
infrastructure} tells signers what happened on the source chain: relayers, RPC
feeds and finality rules.

\paragraph{Deployments.}
A token bridged to several chains has its own bridge \emph{deployment} on
each, and a transfer crosses a \emph{corridor} between two deployments. If all
deployments share one backing, any corridor that can mint or release against
it can drain it.

\paragraph{Containment.}
Even a well-secured bridge can fail; the controls that then limit the damage
are its \emph{containment}. A \emph{rate limit} lets only a fixed amount leave
per period, and an \emph{emergency stop} halts all releases. If the attacker
cannot remove either, it gets only what the rate limit lets through before the
stop: with a \$1M allowance that refills at \$1M a day and a stop after one
hour, about \$1.04M.

\subsection{Security modeling}
A \emph{rubric} rates a system on several properties, each on a scale from
weak to strong. For a bridge, the scale is easy to see: an upgrade key held by
one EOA can drain a bridge in seconds, while a majority multisig behind a
48-hour timelock gives defenders two days to react.

\emph{Attack-defense trees} model how an attack succeeds and where defenses
stop it~\cite{kordy2011foundations}. The root is the attacker's goal, such as
draining an escrow; its children are the steps the attack needs, combined with
AND or OR, and each defense attaches to the step it blocks. Timed variants
record when each step and defense takes effect, so a delay or an emergency stop
can act before the attacker finishes~\cite{gadyatskaya2016timed}.

\subsection{Attacker model}
The attacker can compromise the people, code, controls or off-chain
infrastructure a bridge relies on, as attackers did in the catalogued
incidents, for example by stealing keys or feeding signers false source-chain
events. A bridge operator may also try to look safer than it is, for example
by claiming that its signers are independent or by writing code that
advertises a protection it does not enforce. We also consider breaks of the
cryptography itself: AI already recovers secrets in reduced lattice
instances~\cite{wenger2022salsa}, and a quantum computer would break discrete
logarithms outright~\cite{shor1994}.

\subsection{Related work and the gaps we fill}
\label{sec:related}
\paragraph{Hack datasets.}
Surveys explain how bridges work, what they trust and where they are attacked~\cite{zamyatin2021sok,augusto2024sok,belchior2021survey,mccorry2021sok,zhang2024sok,notland2024sok},
and review bridge and DeFi
hacks~\cite{lee2023sok,zhou2023sok}. Public registries list incidents and
losses~\cite{rekt,defillama,slowmist,chainalysis2022}, and Wu et
al.~\cite{wu2025safeguarding} analyze 49 attacks through September 2024.
\textit{Gap:} none ties each loss to the dependency whose failure caused it, so
none links past hacks to checks of live bridges. Our catalogue does, for
\CorpusN{} incidents through July 2026.

\paragraph{Bridge assessment.}
The Crosschain Risk Framework~\cite{crosschainrisk} scores consensus,
architecture, implementation and operational risk and rates a bridge by its
worst category. L2BEAT~\cite{l2beat} tiers bridges and rollups by design
trust, DeFiSafety~\cite{defisafety} scores process quality, Quantstamp's
framework~\cite{quantstamp_l2} assesses rollup design and operations, and
contract standards~\cite{eea_ethtrust,owasp_scsvs} address code quality.
\textit{Gap:} these rate a bridge from its design or documents, score one
deployment at a time and do not test their scales against losses. Our rubric
decides each check from on-chain evidence, scores an asset across every
deployment that shares its backing, and is tested against past hacks.

\paragraph{Detection and containment.}
Security tools find bugs and attacks~\cite{chaliasos2024tools} from runtime
anomalies~\cite{zhang2022xscope,xchainwatcher2025}, static
analysis~\cite{liao2024smartaxe} and transaction
graphs~\cite{lin2025bridgeshield}. Other designs replace or collateralize the
trusted committee~\cite{zkbridge2022,xclaim} or bound sore-loser
risk~\cite{xue2021soreloser}, and Liu et al.~\cite{liu2024montecrypto} enforce
an accounting invariant inline. \textit{Gap:} none measures how much rate
limits, an emergency stop and a delay on changes cut the loss of real hacks.
Our counterfactual model replays every hack under these levers.

\paragraph{Agentic security analysis.}
Agents plan and run intrusion
steps~\cite{deng2024pentestgpt,shen2025pentestagent,wu2025autopt}, check
model-generated inputs by execution~\cite{yang2025hlpfuzz} and audit contract
code~\cite{sun2024gptscan,liu2025propertygpt}; confinement and benchmarks
limit and measure prompt injection~\cite{debenedetti2025camel,debenedetti2024agentdojo}.
\textit{Gap:} these look for code defects or attacks, not for how a deployed
bridge is configured, and their findings are not tied to recorded evidence.
Our agentic analyzer checks deployed configuration, binds every fact to a
recorded read, scores on the host and repeats unchanged assessments without
the model.
\end{clarityblock}

\section{Design}
\label{sec:design}
\clarity{We start from the bridges that were hacked.
Section~\ref{sec:catalogue} collects the hacks. Section~\ref{sec:rootcause}
finds what failed in each. Section~\ref{sec:damage} finds what limits the
loss once an attack succeeds. Section~\ref{sec:rubric} fuses the two into a
rubric that quantifies the security of live bridges: our bridge security
assessment rubric.}

\subsection{Past hacks}
\label{sec:catalogue}
\clarity{We collected \CorpusN{} bridge hacks from June 2021 to July 2026,
with \$\CorpusGrossB{}B in reported losses. AI agents gathered reports from
hack databases~\cite{rekt,defillama} and security
firms~\cite{slowmist,chainalysis2022}, then searched again for hacks they
missed. A second pass rechecked each amount and cause against the original
sources, and we verified every entry. A hack counts if attackers took funds
through a bridge, including through its website. We also count
\CorpusNearMissN{} cases where a flaw was found or an attack stopped before
anyone lost funds, at \$0. Most hacks
hit Ethereum; others hit Bitcoin, Solana, BNB Chain, Cosmos chains, TON and
Cardano. Appendix~\ref{app:catalogue} lists every hack with its sources.}

\subsection{Root Cause Analysis}
\label{sec:rootcause}
\begin{table}[tp]
\centering
\footnotesize
\setlength{\tabcolsep}{2.5pt}
\renewcommand{\arraystretch}{0.95}
\begin{corrected}
\caption{Each kind of posture failure recurs across years. Columns 21 to 26
 count hacks per year (2026: January to July); code bugs form one row. Losses are reported gross in \$M, rounded, so subtotals need
 not sum.}
\label{tab:taxonomy}
\begin{tabular}{@{}l cccccc r r@{}}
\toprule
\textbf{Class} & \textbf{21} & \textbf{22} & \textbf{23} & \textbf{24} & \textbf{25} &
\textbf{26} & \textbf{n} & \textbf{\$M} \\
\midrule
\multicolumn{9}{@{}l}{\textbf{Posture}}\\
Quorum compromise & 0 & 3 & 4 & 0 & 3 & 1 & 11 & 1{,}095.0 \\
Attester forgery & 3 & 0 & 0 & 0 & 0 & 4 & 7 & 317.6 \\
Admin-key compromise & 0 & 2 & 1 & 1 & 4 & 2 & 10 & 175.7 \\
Frontend hijack & 0 & 1 & 0 & 0 & 0 & 0 & 1 & 0.2 \\
\addlinespace[1pt]
\textit{Subtotal} & \textit{3} & \textit{6} & \textit{5} & \textit{1} & \textit{7} & \textit{7} & \textit{29} & \textit{1{,}588.5} \\
\midrule
\textbf{Code bugs} & 5 & 13 & 2 & 8 & 0 & 15 & 43 & 1{,}933.1 \\
\midrule
\textbf{Total} & \textbf{8} & \textbf{19} & \textbf{7} & \textbf{9} & \textbf{7} & \textbf{22} & \textbf{72} & \textbf{3{,}521.6} \\
\bottomrule
\end{tabular}
\end{corrected}
\end{table}

\paragraph{\clarity{Posture or code.}}
\clarity{We sort each hack with one test: whether the bridge's code did what it
was written to do. If not, the hack is a \emph{code} hack: a bug let the attacker
do something the code should have refused. If it did, the hack is a
\emph{posture} failure: the code worked, and the attacker beat how the bridge
was set up. Two hacks show the difference. At Wormhole, every signer key stayed
safe, but a bug in signature checking let the attacker mint funds without
valid signatures~\cite{pm-wormhole}. At Ronin, the contract worked exactly as
written: it released funds on five valid signatures out of nine, and the
attacker held five keys~\cite{pm-ronin}.}

\clarity{Posture covers who holds the keys, how many signers must approve a
release, whom the bridge trusts to report the other chain, and which limits
stop a release. For software the bridge runs off chain, the line is this: a
bug in signing software is code, while a relayer that accepts a forged event
is posture, because which events count is part of the setup. A posture failure
has no bug for a code audit to find, and most posture weaknesses can be read
from a deployed bridge before an attack (Section~\ref{sec:preexploit}).
We study posture failures, and keep code hacks only to test whether
containment also limits their losses (Section~\ref{sec:damage}).}

\clarity{Posture failures now cause most losses, and they repeat. Since 2023
they caused \CorpusRecentPct{}\% of reported losses, and every one
of the \CorpusRecentPostureN{} posture failures since then repeats a kind seen
by 2022 (Table~\ref{tab:taxonomy}).}
\label{sec:rq1}

\clarity{Posture failures fall into four groups:}
\begin{itemize}[noitemsep,topsep=2pt,leftmargin=1.4em]
\item \clarity{\textbf{People \& Power}: who holds the keys that upgrade the
  bridge, move its funds or replace its signers. \CorpusClassAdminN{} hacks
  stole such a key or misused it from inside (\$\CorpusClassAdminM{}M).}
\item \clarity{\textbf{Code \& Config}: how many signers must approve a
  release, whether they are independent, who may mint, and what backs each
  token. \CorpusClassQuorumN{} hacks took enough signer keys to approve a
  release; they are the costliest kind (\$\CorpusClassQuorumM{}M).}
\item \clarity{\textbf{Off-chain Infrastructure}: how signers learn what
  happened on the other chain. \CorpusClassAttesterN{} hacks fed the signers a
  forged event (\$\CorpusClassAttesterM{}M).}
\item \clarity{\textbf{Website}: one hack hijacked a bridge's website
  (\$\CorpusClassOtherM{}M).}
\end{itemize}

\clarity{The weakest part of a token can sit on a chain its holder never
uses. An assessment must therefore check every
deployment of a token, on every chain.}
\label{sec:rq3}

\subsection{Containment Levers}
\label{sec:damage}
\clarity{Checking a bridge for these root causes can stop known attacks. A
code bug or a new kind of attack still gets through. Containment limits what
such an attack takes, but only if the attacker cannot switch it off.}

\clarity{A rate limit caps how much can leave per period, across every path
that releases funds. Suppose every release draws on one allowance that starts
at $C_0$ and refills by at most $rt$ over any interval $t$. Then}
\setlength{\abovedisplayskip}{3pt}\setlength{\belowdisplayskip}{3pt}%
\begin{equation}\label{eq:bound}
\mathrm{loss} \;\leq\; \min\{\,V,\; C_0 + r\Delta\,\},
\end{equation}
\clarity{where $V$ is the value at risk, $r$ the refill rate and $\Delta$ the
time to detect the attack and stop the bridge. If the attacker can lift the
limit, the bound is $V$.}

\clarity{An emergency stop halts all releases, but it protects only what is
still there. The rate limit buys the time to use it. In our loss model, a rate
limit and a stop at every place funds leave, with everything else unchanged,
cut losses from \$\CorpusGrossB{}B to \$\RateStopResidualM{}M,
\RateStopFold{}-fold. Ronin
and KelpDAO each fall to \$1.04M (Section~\ref{sec:counterfactual}).}

\clarity{A privileged change, such as a new owner or a new set of signers,
often comes just before a drain. In \CorpusTimingWithinHourN{} of the
\CorpusTimingN{} hacks with known timing, the drain followed within an hour
(Figure~\ref{fig:drain}). An alert alone comes too late. A delay that every
such change must wait out gives defenders time to use the stop.}

\clarity{Fixing the root causes of Section~\ref{sec:rootcause} and adding
these levers together cut losses \IdealControlReductionFold{}-fold, to
\$\IdealControlResidualM{}M (Section~\ref{sec:counterfactual}).}

\definecolor{BBchain}{HTML}{134A86}
\definecolor{BBrep}{HTML}{EB6834}
\definecolor{BBrule}{HTML}{9BA3AE}
\providecommand{\drainrow}[3]{\makebox[39pt][l]{#1}%
  \makebox[30pt][r]{\scriptsize #2}\makebox[25pt][r]{\scriptsize #3}}
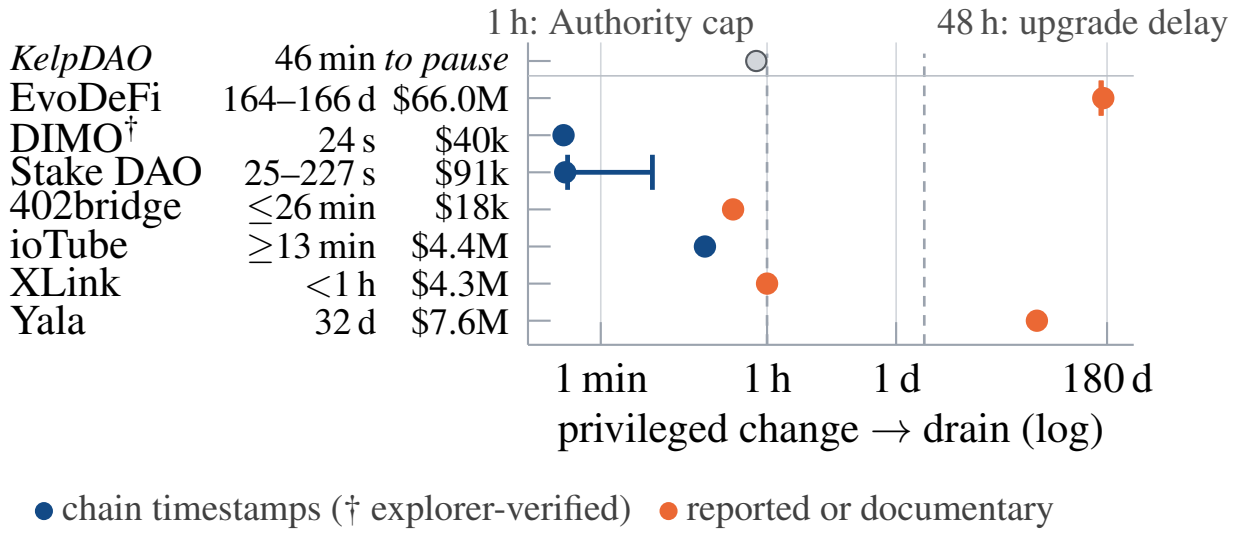
\begin{figure}[tp]
\centering
\begin{tikzpicture}
\begin{axis}[
  scale only axis=true, width=114pt, height=2.0cm,
  xmode=log, log basis x=10,
  xmin=10, xmax=3e7,
  ymin=0.35, ymax=8.5,
  xtick={60,3600,86400,15552000},
  xticklabels={1\,min,1\,h,1\,d,180\,d},
  xticklabel style={font=\footnotesize},
  xlabel={privileged change $\to$ drain (log)},
  xlabel style={font=\footnotesize, yshift=3pt},
  ytick={1,2,3,4,5,6,7,8},
  yticklabels={{\drainrow{Yala}{32\,d}{\$7.6M}},
    {\drainrow{XLink}{$<$1\,h}{\$4.3M}},
    {\drainrow{ioTube}{$\ge$13\,min}{\$4.4M}},
    {\drainrow{402bridge}{$\le$26\,min}{\$18k}},
    {\drainrow{Stake DAO}{25--227\,s}{\$91k}},
    {\drainrow{DIMO$^{\dagger}$}{24\,s}{\$40k}},
    {\drainrow{EvoDeFi}{164--166\,d}{\$66.0M}},
    {\drainrow{\scriptsize\textit{KelpDAO}}{46\,min}{\scriptsize\textit{to pause}}}},
  yticklabel style={font=\footnotesize},
  axis lines*=left,
  axis line style={draw=BBrule, line width=0.4pt},
  tick style={draw=BBrule, line width=0.4pt},
  xmajorgrids, grid style={draw=BBrule!45, line width=0.3pt},
  clip=false,
]
\draw[BBrule, line width=0.5pt, dashed, dash pattern=on 2pt off 1.6pt]
  (axis cs:3600,0.35) -- (axis cs:3600,8.2);
\draw[BBrule, line width=0.5pt, dashed, dash pattern=on 2pt off 1.6pt]
  (axis cs:172800,0.35) -- (axis cs:172800,8.2);
\node[font=\scriptsize, anchor=south east, inner sep=2pt, text=black!70]
  at (axis cs:3480,8.2) {1\,h: Authority cap};
\node[font=\scriptsize, anchor=south west, inner sep=2pt, text=black!70]
  at (axis cs:179000,8.2) {48\,h: upgrade delay};
\addplot[only marks, mark=*, mark size=1.9pt, draw=BBrep, fill=BBrep] coordinates {(2764800,1)};
\addplot[only marks, mark=*, mark size=1.9pt, draw=BBrep, fill=BBrep] coordinates {(3600,2)};
\addplot[only marks, mark=*, mark size=1.9pt, draw=BBchain, fill=BBchain] coordinates {(780,3)};
\addplot[only marks, mark=*, mark size=1.9pt, draw=BBrep, fill=BBrep] coordinates {(1560,4)};
\draw[BBchain, line width=0.9pt, |-|] (axis cs:25,5) -- (axis cs:227,5);
\addplot[only marks, mark=*, mark size=1.9pt, draw=BBchain, fill=BBchain] coordinates {(25,5)};
\addplot[only marks, mark=*, mark size=1.9pt, draw=BBchain, fill=BBchain] coordinates {(24,6)};
\draw[BBrep, line width=0.9pt, |-|] (axis cs:14169600,7) -- (axis cs:14342400,7);
\addplot[only marks, mark=*, mark size=1.9pt, draw=BBrep, fill=BBrep] coordinates {(14169600,7)};
\draw[BBrule!70, line width=0.3pt] (axis cs:10,7.6) -- (axis cs:3e7,7.6);
\addplot[only marks, mark=*, mark size=1.9pt, draw=BBrule!55!black, fill=BBrule!45] coordinates {(2760,8)};
\node[font=\scriptsize, anchor=north west, inner sep=1pt, text=black!75]
  at (rel axis cs:0,0) [yshift=-27pt, xshift=-94pt]
  {\tikz{\fill[BBchain] (0,0) circle (1.7pt);}~chain timestamps
   ($\dagger$~explorer-verified)\ \ \
   \tikz{\fill[BBrep] (0,0) circle (1.7pt);}~reported or documentary};
\end{axis}
\end{tikzpicture}
\caption{Most drains with known timing followed the privileged change within an hour: \CorpusTimingWithinHourN{} of \CorpusTimingN{}. Each row places one incident on a log scale from change to drain; blue marks come from chain timestamps. Dashed lines mark the Authority cap's one-hour delay and the upgrade delay's 48-hour top tier. The top row shows KelpDAO's time from attack to pause.}
\label{fig:drain}
\end{figure}

\subsection{The bridge security assessment rubric}
\label{sec:rubric}
\clarity{We fuse the root causes and the levers into one rubric: each becomes
a question whose tiers score a live bridge from weak to strong. We first
follow one hack from root cause to score. We then describe the
questions, the hard caps, and how a token is scored across chains.}

\clarity{Take KelpDAO. Its bridge accepted a message once a single verifier
approved it. The matching question asks how many signers must approve a
release. A single signer, or any threshold of half or less, is tier 1. At
least two thirds, with no one able to bypass them, is tier 5. One approver is
fatal however strong the rest of the bridge is, so this setting fires a hard
cap and the bridge scores 1. Our check of chain data from before the attack
finds this setting (Section~\ref{sec:preexploit}). The missing lever shows too:
in our loss model, a rate limit and a stop hold the \$292M loss to \$1.04M
(Section~\ref{sec:counterfactual}).}

\paragraph{Questions.}
\clarity{Our rubric asks 26 weighted questions in four categories
(Table~\ref{tab:rubric}). People \& Power, Code \& Config and Off-chain
Infrastructure are the first three groups of Section~\ref{sec:rootcause} and
hold the root causes. Controls \& History
holds the levers of Section~\ref{sec:damage}, the website and the bridge's
incident history. The categories weigh 40, 35, 25 and 15. Each question scores
a deployment from 1 (weak) to 5 (strong), using on-chain reads or public
evidence. Four more questions (Q27--Q30) ask what an attacker who breaks the
cryptography could do; they change no score. Each question also names the
on-chain change that should trigger a new assessment
(Section~\ref{sec:monitoring}).}

\newlength{\rubricQw}\newlength{\rubricCw}
\newlength{\rubricPw}\newlength{\rubricSw}\newlength{\rubricSCw}
\newcommand{\rubriccols}[1]{%
  \setlength{\tabcolsep}{3pt}%
  \setlength{\rubricQw}{2.45cm}%
  \setlength{\rubricCw}{1.40cm}%
  \setlength{\rubricPw}{\dimexpr\textwidth-6\tabcolsep-\rubricQw-\rubricCw\relax}%
  \setlength{\rubricSw}{\rubricPw}%
  \setlength{\rubricPw}{#1\rubricPw}%
  \setlength{\rubricSw}{\dimexpr\rubricSw-\rubricPw\relax}%
  \setlength{\rubricSCw}{\dimexpr\rubricSw+\rubricCw+2\tabcolsep\relax}%
}
\newcolumntype{R}[1]{>{\raggedright\arraybackslash}p{#1}}
\newcommand{\nocap}[1]{\multicolumn{2}{R{\rubricSCw}@{}}{#1}}

\newlength{\rubricMQw}\newlength{\rubricMTw}
\newcommand{\rubricmergedcols}[1]{%
  \setlength{\tabcolsep}{3pt}%
  \setlength{\rubricMTw}{\dimexpr\textwidth-2\tabcolsep\relax}%
  \setlength{\rubricMQw}{#1\rubricMTw}%
  \setlength{\rubricMTw}{\dimexpr\rubricMTw-\rubricMQw\relax}%
}

\begin{table*}[!t]
\centering\scriptsize\linespread{0.96}\selectfont
\rubricmergedcols{0.507}\renewcommand{\arraystretch}{1}
\caption{\correction{The rubric, Q1--Q14: each row gives the property, read and score tiers. Brackets name hard caps; shaded rows are questions our analyzers do not score, as they need off-chain data. Version 2026-10-25.1. Ethereum Improvement Proposal (EIP); decentralized autonomous organization (DAO); Universal Upgradeable Proxy Standard (UUPS).}}
\label{tab:rubric}
\begin{tabular}{@{}R{\rubricMQw} R{\rubricMTw}@{}}
\toprule
\textbf{Question, property and read} & \textbf{Score tiers} \\
\midrule
\addlinespace[1pt]
\multicolumn{2}{@{}l}{\textbf{People \& Power (weight 40)}}\\
\textbf{Q1} Upgrade-authority composition [Authority]. Independent parties needed to upgrade in-scope code. \texttt{owner()} and EIP-1967 admin slot on each canonical contract and ProxyAdmin; weakest path as EOA / Safe$(m,n)$ / DAO; shared signers. \emph{Monitor:} \texttt{OwnershipTransferred}, \texttt{AdminChanged}, Safe owner/threshold change. & \tiers{\tier{5}{owner is a timelock or on-chain DAO-governor; or immutable \emph{and} no owner role over remote counterparts, verification modules or verifier sets (immutable code alone does not remove configuration authority); none shared}\tier{3}{majority-threshold Safe ($m \geq \lfloor n/2 \rfloor + 1$, $n \geq 3$), except 2-of-3, no timelock}\tier{2}{sub-majority Safe, 2-of-3 Safe, or $>$50\% shared signers}\tier{1}{EOA, or any Safe whose threshold is 1 (one key suffices, whatever the signer count)}} \\
\textbf{Q2} Upgrade delay [Authority]. Min delay on the upgrade path: \texttt{getMinDelay()} or equivalent; guardian early-execute; timelock admin also proposer; EOA owner $=$ no delay. \artifact{\emph{Throughout this rubric a control is \textbf{timelocked} only where its delay is defender-actionable, meaning $\geq$1\,h, the minimum for generic timelocked-authority credit; a shorter delay is scored as none wherever a tier says ``timelocked'' without naming a duration.}} & \tiers{\tier{5}{proved no code-replacement path, or $\geq$48\,h with no bypass}\tier{4}{24--48\,h}\tier{3}{1--24\,h}\tier{2}{positive delay $<$1\,h, or guardian early-execute}\tier{1}{none}} \\
\textbf{Q3} Verifier-set control [Quorum]. Who can replace the verification path? Read the authorities that can change verifier sets, verification modules or trusted remote contracts; their approval requirements and enforced delays; and whether routing is fixed by the application or inherited from a provider. \emph{Monitor:} changes to those authorities and configurations. & \tiers{\tier{5}{immutable, or $\geq$48\,h-timelocked multisig, routing pinned}\tier{3}{timelocked, or immutable AND-composition across independent providers, none forging alone}\tier{1}{one untimelocked party can swap, or one provider forges alone}}. Pinned routing counts only if the pinning party cannot unpin it: an untimelocked authority able to replace the verification path scores 1 regardless of pinning. \\
\textbf{Q4} Custody-movement roles [Authority]. Roles that move or seize custody with no inbound message (rescue, sweep, \texttt{withdrawTo}, \texttt{take}); holder type, timelock. Scoped to the bridge's own roles over what it escrows or mints; the bridged asset's issuer powers (\texttt{blacklist}, \texttt{freeze}, issuer \texttt{burn}) hold on every chain the asset exists on and are not bridge posture. Throttle and pause are liveness controls, scored at Q15 and Q17, not here; the bridge's own \texttt{mint}/\texttt{burn} is its minting right, scored at Q8. & \tiers{\tier{5}{none}\tier{3}{all timelocked and multisig-held}\tier{2}{one untimelocked, not EOA-held}\tier{1}{any EOA-held, or multiple untimelocked}} \\
\rowcolor{EvidenceShade}
\textbf{Q5} Key-rotation history (90\,d). Unexplained or threshold-lowering signer change in 90\,d. Owner add/remove, threshold-change, role grant/revoke events on in-scope contracts. \artifact{An absent explanation is unexplained; a present but unverifiable explanation is distinct. A cut requires an actual threshold decrease, not merely a currently weak threshold.} & \tiers{\tier{5}{none, or all documented and justified}\tier{3}{present justification unverifiable}\tier{2}{any unexplained or affirmatively unjustified change}\tier{1}{threshold cut below strict majority}} \\
\addlinespace[1pt]
\multicolumn{2}{@{}l}{\textbf{Code \& Config (weight 35)}}\\
\textbf{Q6} Release threshold [Quorum]. $M$-of-$N$ on the fund-releasing path\artifact{ (NTT \texttt{threshold()}, LZ ULN required DVNs, CCIP RMN, Hyperlane ISM \texttt{threshold()})}; bypass selectors. Read the \emph{inbound} configuration for the source route\artifact{ (LZ \texttt{getReceiveLibrary} plus its receive ULN)}, not the send side; where the two diverge, the inbound one scores. Classify the holder of any release-bypassing or threshold-rotating role as EOA / multisig / timelock. \emph{Monitor:} drop in release threshold\artifact{ or \texttt{requiredDVNCount}}, and any change to the release-path library or verifier, including a disabled path becoming live. & \tiers{\tier{5}{$\geq2/3$, no single-party bypass}\tier{3}{strict majority below $2/3$}\tier{1}{at or below half (e.g.\ 2-of-4), $1/1$\artifact{ (incl.\ \texttt{requiredDVNCount}\,$=$\,1)}, threshold not gated on release, or a bypass one party can exercise with no defender-actionable delay}}. A path disabled rather than thresholded (blocked or unset library, paused receive) scores \textbf{1} unless disabling is irreversible or re-enabling is delayed: a single untimelocked re-enabler is the effective threshold. Rotation authority held by a multisig or timelock scores at Q3. \\
\textbf{Q7} Verifier independence [Quorum]. Do verifiers share an organization, infrastructure account, upstream provider or source-chain client? Use chain state where it exposes those relationships and first-party operator attestations otherwise, never a curated roster. Distinct addresses do not establish independence. Score the worst axis; unresolved identities receive tier 2 and leave their Quorum contribution \emph{unknown}. & \tiers{\tier{5}{four axes distinct}\tier{3}{majority independent on the worst axis}\tier{2}{majority share one axis, or single attester}\tier{1}{majority shared on $\geq$2 axes}} \\
\textbf{Q8} Mint authority [Backing]. Mint off the verified-release path. Minter set\artifact{ (NTT \texttt{minter()}, FiatToken \texttt{isMinter}, CCIP \texttt{MINTER\_ROLE}, xERC20 \texttt{mintingMaxLimitOf})} vs its trusted-mint manifest; each message-gated? \artifact{An owner read alone does not enumerate minters; a failed direct mint call does not establish a release gate. Off-manifest and ungated are independent failures.} \emph{Monitor:} minter grant/revoke (\texttt{MinterAdded}, \texttt{RoleGranted}); ungated mint. & \tiers{\tier{5}{every minter is within the manifest and message-gated, and at least one minter is a contract whose release gate the minter set cannot itself relax (external threshold, per-bridge limit, or immutable gate)}\tier{3}{the minter set cannot be read; or it is readable, in-manifest and message-gated, but no minter's gate lies beyond the set's own reach (as when it is only the bridge's own contract, whose gate reduces to that contract's upgrade authority)}\tier{1}{off-manifest or ungated minter}} \\
\textbf{Q9} Backing anchor [Backing]. Demonstrated locking anchor for every minted asset (minted $\leq$ locked). Tier~5 requires a demonstrated in-corridor anchor and coherent topology, without requiring complete numerical reconciliation across chains; it does not certify numerical reserve adequacy. Establish minting applicability first; unread modes without an established minting subject withhold the score. \artifact{Resolve code and state separately for each chain instance; matching addresses do not establish matching modes or reserves.} Lock-mint: lock-side escrow vs mint-side \texttt{totalSupply()} corridor-wide; burn-mint: canonical issuer. The anchor must be resolvable \emph{in-corridor}; a locking escrow, or an in-corridor canonical issuer whose reserve is readable; an off-corridor issuer whose reserve is not statically resolvable from the corridor is tier~3, not tier~5. \emph{Monitor:} \texttt{totalSupply()} vs locked each block; alert on excess. & \tiers{\tier{5}{anchored, topology coherent}\tier{3}{minting established; corridor-wide backing vantage not statically resolvable}\tier{1}{wrapped asset, no locking anchor}} \\
\textbf{Q10} Remote counterpart registration. Does each receiving contract accept messages only from its expected remote counterparts? Compare configured chain/domain identifiers and contract addresses with the trusted deployment manifest; an absent required entry is missing. \emph{Monitor:} changes to remote addresses and route registrations. & \tiers{\tier{5}{all present, matching, none unexpected}\tier{2}{unexpected but plausible counterpart awaiting review}\tier{1}{expected counterpart missing, or counterpart definitively unverified by its Q12 publication panel}} \\
\textbf{Q11} Code mutability. Three EIP-1967 slots (implementation, admin, beacon); resulting pattern. \artifact{Missing runtime or slot reads withholds classification; a known standard proxy with a partially resolved implementation is tier 3. An unrecognized pattern after complete successful investigation is distinct from unread evidence.} & \tiers{\tier{5}{proved immutable, slots \texttt{0x0}, no replacement or delegation path}\tier{4}{Transparent/UUPS/Beacon proxy, verified impl.}\tier{3}{same, slots or implementation verification partially resolvable}\tier{2}{custom or Diamond dispatcher}\tier{1}{unrecognized after complete investigation}} \\
\textbf{Q12} Source verification [Authenticity]. Deployed bytecode matching published source, every in-scope contract. Per declared implementation, proxy and routing component on both chains: publication-panel status and bytecode match. \artifact{Score each contract first; an unresolved contract with no positive panel result withholds the deployment score unless a known violation already proves tier 1.} & \tiers{\tier{5}{verified on every panel member, bytecode match}\tier{3}{every contract verified on some panel members, not all}\tier{1}{unverified everywhere, or any bytecode mismatch}} \\
\rowcolor{EvidenceShade}
\textbf{Q13} Audit assurance. Qualified review of this exact deployed code, findings remediated. \texttt{keccak256} of runtime bytecode vs recompiled audited commit; open high/critical findings; scope; days since audit; auditor tier. & \tiers{\tier{5}{exact bytecode, full scope, none open, $<$365\,d, tier-1}\tier{4}{as 5 but 365--730\,d}\tier{3}{$>$730\,d, low-severity open finding, or non-critical surface unscoped}\tier{2}{bytecode diverged, critical surface unscoped, or tier-3 auditor}\tier{1}{no audit}} \\
\textbf{Q14} Replay protection. One cross-chain message processed twice. Verified receive path\artifact{ (NTT \texttt{isMessageExecuted}, LZ inbound nonce, CCIP off-ramp state)}: executed-state tracking, ordering, identifier. \artifact{A canonical getter is discovery only; getters, declarations and unreachable byte markers do not prove control flow. Unread enforcement withholds the score.} & \tiers{\tier{5}{checked before the state write, keyed on full identifier (chain, sequence, payload hash), reverts on duplicate}\tier{3}{demonstrated guard, non-standard (partial identifier, silent no-op), or incomplete assurance about a demonstrated guard}\tier{1}{absent, or after the state change}} \\
\bottomrule
\end{tabular}
\end{table*}

\begin{table*}[!t]\ContinuedFloat
\centering\scriptsize\linespread{0.96}\selectfont
\rubricmergedcols{0.507}\renewcommand{\arraystretch}{1}
\caption{\correction{The rubric (continued), Q15--Q30. Q27--Q30 are unweighted and change neither posture scores nor caps. Domain Name System Security Extensions (DNSSEC); Hypertext Transfer Protocol (HTTP) Strict Transport Security (HSTS); Subresource Integrity (SRI); application programming interface (API).}}
\label{tab:rubric-b}
\label{tab:rubric-pq}
\begin{tabular}{@{}R{\rubricMQw} R{\rubricMTw}@{}}
\toprule
\textbf{Question, property and read} & \textbf{Score tiers} \\
\midrule
\addlinespace[1pt]
\multicolumn{2}{@{}l}{\textbf{Controls \& History (weight 25)}}\\
\textbf{Q15} Rate limits. Drain per window. Transfer or mint limit and refill duration\artifact{ (NTT \texttt{inboundLimit}/\texttt{rateLimitDuration}, CCIP capacity/refill, Hyperlane \texttt{RateLimitHook})}, normalized to a maximum executable drain over 24\,h, including immediately available bucket capacity, refill and queued release paths under the enforced semantics. & \tiers{\tier{5}{$\leq$5\% of backing per 24\,h}\tier{4}{$\leq$20\%}\tier{3}{$\leq$50\%}\tier{2}{$>$50\% but below 100\% of backing, or an enforced ceiling with no refill window}\tier{1}{no limit enforced, or a ceiling at or above 100\% of backing (including a limit set to the type maximum): a check that executes but cannot bind on any window is not an enforced limit}} \\
\textbf{Q16} Destination supply cap. Mintable supply capped regardless of bridge flow. Mint-side cap\artifact{ (\texttt{cap()}/\texttt{maxSupply()}/CCIP lane cap)} vs \texttt{totalSupply()}; applies to each minting subject, including the mint destination of a lock-mint corridor. A release-only subject is N/A. A getter alone does not prove a supply cap; compare with the same enforced 24\,h flow measure as Q15. \artifact{A demonstrated cap with a proved absent finite flow baseline scores 3; an unread necessary baseline withholds unless an independent bound decides the tier.} & \tiers{\tier{5}{cap with headroom $\leq5\times$ the Q15 24\,h limit}\tier{3}{loose cap, or no Q15 limit to compare}\tier{2}{no cap}} \\
\textbf{Q17} Pause / circuit-breaker. Defenders able to halt an active attack. Pause role, holder type, latency, scope, state; scan for an automatic breaker (risk-network halt, cumulative-drain trip). \artifact{Bounds require executed auto-resume or user-exit logic; a timestamp getter alone proves neither bounded nor indefinite pause.} & \tiers{\tier{5}{self-controlled multisig pause with redundant signers ($2\leq m<n$), or an automatic no-signature breaker on the affected lane}\tier{3}{EOA or 1-of-1 pauser; unanimous multisig ($m=n>1$); EOA-owned pause module; or a dedicated lane breaker gated by an external quorum}\tier{2}{deployment-wide halt requiring an external transfer-committing quorum}\tier{1}{no pause or breaker on the release/mint path, or only an underlying layer's global governance halt}}. Never-tested manual pause caps at \textbf{4}; unbounded pause with no auto-resume or user exit caps at \textbf{3}, an indefinite freeze being itself a denial of withdrawal. \\
\rowcolor{EvidenceShade}
\textbf{Q18} Public monitoring. Users and integrators, not just the operator, able to detect an incident. Dashboard, status/health endpoint, alert feed, from official site and docs. & \tiers{\tier{5}{all three}\tier{4}{two}\tier{3}{one}\tier{1}{none}} \\
\rowcolor{EvidenceShade}
\textbf{Q19} Technology incident history. This technology exploited at the contract or ingestion layer. Rekt, DefiLlama, status pages, auditor advisories; time since the worst-layer one. & \tiers{\tier{5}{none}\tier{4}{decayed contract exploit ($\geq$3\,y, no recurrence), or no public ingestion post-mortem}\tier{3}{resolved exploit $>$2\,y ago}\tier{2}{exploit $<$2\,y ago, or integrity event}\tier{1}{$<$6-month exploit with loss in this family}} \\
\rowcolor{EvidenceShade}
\textbf{Q20} Deployment incident history. Incident or emergency action on this deployment. \texttt{Paused}/\texttt{Upgraded}/governance events over its lifetime, joined to an incident database; emergency $=$ inside the Q2 timelock window, or pause-then-upgrade. Event age does not establish resolution or non-emergency status; an old unadjudicated action is unknown. & \tiers{\tier{5}{no emergency actions}\tier{4}{single affirmatively resolved non-emergency event}\tier{2}{one proved emergency action within 90\,d or affirmatively unresolved}\tier{1}{multiple}} \\
\rowcolor{EvidenceShade}
\textbf{Q21} Frontend / DNS. User doing everything right still rugged via the UI. DNSSEC, registrar lock, TLS/HSTS, UI-advertised vs canonical addresses (Q1), SRI on cross-origin scripts, active typosquats. & \tiers{\tier{5}{DNSSEC, registrar lock, SRI, and the UI address pinned to canonical: no single unlocked party can repoint DNS or swap the served bundle}\tier{4}{valid TLS and HSTS, address matching, registrar lock and SRI present, but DNSSEC missing (repoint requires defeating a protected path)}\tier{2}{DNS or registrar mutable with no lock, or no SRI: one party repoints the origin or swaps scripts with no delay, whether or not an attack is live}\tier{1}{no DNSSEC, no lock and no SRI, or an observed live address mismatch or typosquat serving the production UI}}. Apply the lowest matching tier; a live mismatch or unprotected script/registrar path cannot earn tier 4 from TLS alone. \\
\addlinespace[1pt]
\multicolumn{2}{@{}l}{\textbf{Off-chain Infrastructure (weight 15)}}\\
\rowcolor{EvidenceShade}
\textbf{Q22} Ingestion tamper-evidence. Third party able to verify the running binary against its source. Source public, build reproducible, binary signed, live runtime attestation bound to it. & \tiers{\tier{5}{open, reproducible, signed, attested}\tier{4}{open and signed}\tier{3}{open but unsigned}\tier{2}{closed but independently audited}\tier{1}{closed, no audit}}. N/A if validity is fully on-chain. \\
\rowcolor{EvidenceShade}
\textbf{Q23} RPC trust posture and event binding [Provenance (at score 1)]. Attester agreement across independent RPCs before signing, fail-closed, crediting only events derived from canonical source state. Off-chain evidence is required: signer-key counts establish neither RPC-provider nor operator-organization diversity. Provider count; payload-level quorum at a finalized block; fail-mode on disagreement or unreachability; credited-event resolution (asset by canonical token address, not symbol; deposit events; emitter verified canonical). & \tiers{\tier{5}{$\geq$3 providers, payload-level quorum, fail-closed, all credited fields from canonical source state}\tier{3}{multi-provider health-check, no payload quorum, or derivation present but not fully verifiable}\tier{2}{failover, no quorum, silent fallback}\tier{1}{single or undocumented, asset resolved by symbol, or a trusted emitter with no canonical bind}} \\
\rowcolor{EvidenceShade}
\textbf{Q24} Per-chain finality. Confirmation rules calibrated to each source chain's reorg risk. Attester's documented rule per chain, by time-to-finality. & \tiers{\tier{5}{per-chain rules matching each profile (Ethereum finalized or $\geq$12 confirmations, Solana finalized, deterministic-finality 1)}\tier{3}{single global rule, or a chain below its reorg-safe depth}\tier{1}{undocumented}} \\
\rowcolor{EvidenceShade}
\textbf{Q25} Reorg / attestation-cancel. Cancel before minting when the source reorgs past what the attester saw. Reorg policy, cancel-or-replace mechanism, finality depth (Q24), pre-finality fast lane, past cancellations. & \tiers{\tier{5}{cancel mechanism exercised in production}\tier{4}{present but unexercised, or deep-finality policy}\tier{3}{no policy but observed deep confirmation}\tier{2}{undocumented with a fast lane}\tier{1}{undocumented}} \\
\rowcolor{EvidenceShade}
\textbf{Q26} Ingestion-health monitoring. Health visible to anyone in real time. Per-message status API, per-attester signing/divergence, RPC liveness, attestation latency. & \tiers{\tier{5}{all four plus an alert feed}\tier{4}{per-message plus per-attester}\tier{3}{per-message only}\tier{2}{coarse status dashboard}\tier{1}{none}} \\
\addlinespace[1pt]
\multicolumn{2}{@{}l}{\textbf{Cryptographic Resilience (unweighted; agentic path)}}\\
\textbf{Q27} Mathematical break. If one hardness assumption behind the keys breaks (elliptic curves, pairings, factoring or lattices), for example through AI-found mathematics, can a privileged action (upgrade, configuration, mint, release, signer rotation, governance vote, unpause) still complete on any chain? A stop counts if no broken key can lift it and the action waits at least 1\,h behind it. & \tiers{\tier{5}{no privileged action completes under any single-assumption break}\tier{3}{some completes, but every such action is guarded by a surviving stop}\tier{1}{some completes with no surviving stop}} \\
\textbf{Q28} Quantum break. The same paths, assuming a quantum computer breaks every scheme the state of the art (NIST post-quantum signature standards, hash-based signatures) does not regard as quantum-resistant. & \tiers{\tier{5}{no privileged action completes}\tier{3}{some completes, but every such action is guarded by a surviving stop}\tier{1}{some completes with no surviving stop}} \\
\textbf{Q29} Bunker mode. A pre-arranged emergency stop that, once engaged, halts every privileged path, including upgrades and role changes that could remove it, and that no key broken under Q27 or Q28 can undo, revoke, disable or bypass. & \tiers{\tier{5}{such a stop exists and halts every privileged path}\tier{3}{it holds against every broken key but halts only some paths}\tier{1}{every stop can be undone, revoked, disabled or bypassed with a broken key, or none exists}} \\
\textbf{Q30} Custom verifier. A privileged path relies on verification code not established, from source or decompiled bytecode, as a valid implementation of a recognised scheme: each signature or proof checked, bound to the authorized signer set or verifying key, distinct signers counted against the threshold. Such a verifier counts as broken in Q27--Q29. & \tiers{\tier{5}{no}\tier{1}{yes}} \\
\bottomrule
\end{tabular}
\end{table*}

\paragraph{Tiers.}
\clarity{Each question's tiers run from the setting behind past hacks to the
setting that blocks them. Tier 1 is what we found at the hacks: an upgrade key
held by one account with no delay, as at DIMO~\cite{pm-dimo}; a single
verifier, as at KelpDAO; a relayer that identifies a token by its symbol, as at
THORChain in 2021~\cite{pm-thorchain-bifrost}. Tier 5 is the setting that stops
that attack: at least two thirds of independent signers, a 48-hour delay on
upgrades, a rate limit of at most 5\% of the backing per day. The tiers between
them are partial fixes, ordered by how much harder they make the attack.
Section~\ref{sec:validation} tests this order: on all \PhysicalTierBranches{}
tier steps that change what an attacker can take, the higher tier loses less in
our loss model and never more, for one setting of each tier on a declared set of attacks. Where each cut-off falls, such as 48 hours or
5\%, is our judgment (Section~\ref{sec:limits}).}

\paragraph{Hard caps.}
\clarity{Some weaknesses are fatal however strong the rest is. A hard cap marks
such a weakness and sets the score to 1. There are five:}
\begin{itemize}[noitemsep,topsep=2pt,leftmargin=1.4em]
\item \textbf{Authority} fires when a weak key can upgrade the bridge or move
  its custody without at least one hour of delay, or through a usable
  bypass. \clarity{A key is weak when one signature suffices, a multisig
  needs at most half its signers or two of three, most of its signers are
  shared, or a custody role has no timelock.}
\item \textbf{Quorum} fires when the evidence proves a release threshold of
  one signer or at most half, signers concentrated in one party, or one party
  that can replace the verifiers without delay.
\item \textbf{Backing} fires when a mint is not limited to the declared
  minters or not gated by a verified message, or when the backing is missing
  or does not add up.
\item \textbf{Authenticity} fires when a contract's source is unverified on
  every explorer we check, or its bytecode differs from the published source.
\item \textbf{Provenance} fires when signers rely on a single or undocumented
  RPC provider, credit an event not bound to the source chain, or keep going
  when a check fails.
\end{itemize}
\clarity{A cap fires on \CorpusCapsN{} of the \CorpusPostureN{} posture hacks:
Quorum on \CorpusQuorumN{}, Authority on \CorpusAuthorityN{}, Backing on
\CorpusBackingN{} and Provenance on \CorpusProvenanceN{}. The remaining one, Celer's
hijacked website, gets the lowest tier of the website question (Q21), which
has no cap: a hijacked website takes what its visitors approve, not what the
bridge holds, and no chain data can check a website. For \CorpusPreExploitCapsN{} of the
\CorpusPostureN{}, chain data from before the attack already shows the weakness
(Sections~\ref{sec:caps} and~\ref{sec:preexploit}).}

\paragraph{From deployment to token.}
\clarity{A token is only as safe as its weakest deployment. On each question, a
token takes the weakest answer among its deployments that can mint it or
release its backing, on any chain. When all deployments share one backing, a failure at any one of
them can empty it. A cap on any deployment drops the token to 1, and the
report names that deployment. If one of them cannot be assessed, the token's
score is withheld. Issuer powers to freeze or burn the token stay outside the
assessment.}

\section{Implementation}
\label{sec:system}
\clarity{In Section~\ref{sec:design}, we built the rubric and its tiers from
past hacks. We now turn the rubric into an active analyzer that reviews
bridges and quantifies their security. This section covers our implementation
of a live analyzer, which we test on the seven bridge stacks that our \EvalBoardCorridors{} tokens use. We cover two sets
of analyzers. A hard-coded bespoke analyzer per bridge reads the rubric's
values to judge and score the bridge (Section~\ref{sec:analyzer}). A generic
agentic analyzer and toolset uses large language models to carry out the same
checks (Section~\ref{sec:agentic}). Since configurations can change, we also
build a live monitoring system (Section~\ref{sec:monitoring}).}
\ifanon\else These tools serve the Monad ecosystem.\fi

\subsection{\clarity{Bespoke Analyzers}}
\label{sec:analyzer}
\clarity{Our bespoke analyzers read each bridge stack's contracts with code
written for that stack. We state their requirements, then their
architecture, how they follow a token across chains and how they watch for
changes.}

\paragraph{\clarity{Requirements.}}
\clarity{Our bespoke analyzers must meet five requirements:}
\begin{itemize}[noitemsep,topsep=2pt,leftmargin=1.4em]
\item \clarity{\textit{Coverage.} They support the seven bridge stacks that our \EvalBoardCorridors{} tokens use: LayerZero V1 and V2~\cite{layerzero}, Chainlink's
  Cross-Chain Interoperability Protocol (CCIP)~\cite{ccip}, Wormhole NTT and
  Portal~\cite{wormhole}, Circle CCTP v2~\cite{cctp} and Hyperlane
  Warp~\cite{hyperlane}.}
\item \clarity{\textit{Scoring.} For each stack, they assign a tier to every
  question that chain data can decide and decide the five caps.}
\item \clarity{\textit{Reproducibility.} The same chain state always gives the
  same report, so anyone can rerun an assessment.}
\item \clarity{\textit{Cross-chain.} They assess a token on every chain it
  reaches, since its weakest deployment can be on a chain its holder never
  uses (Section~\ref{sec:rootcause}).}
\item \clarity{\textit{Monitoring.} They notice when a bridge's configuration
  changes, since attackers change it shortly before a drain
  (Section~\ref{sec:damage}).}
\end{itemize}

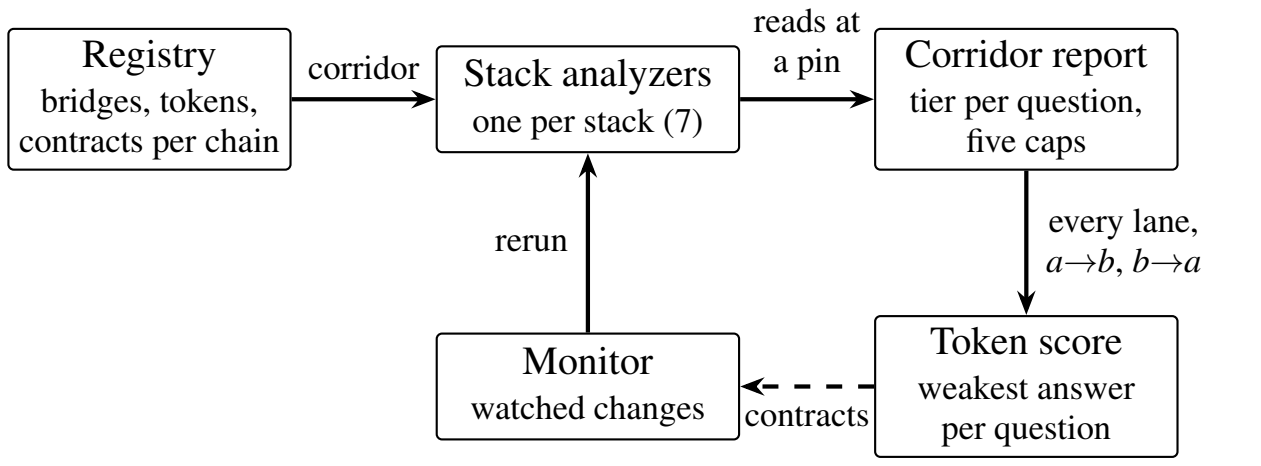
\begin{figure}[tp]
\centering
\def\an#1{{\fontsize{6.5}{7.5}\selectfont #1}}
\begin{tikzpicture}[
  x=1mm, y=1mm,
  box/.style={draw, rounded corners=1pt, align=center, inner sep=2pt,
    font=\fontsize{7.5}{8.5}\selectfont, minimum height=7mm},
  arr/.style={-{Stealth[length=1.6mm]}, thick},
  lbl/.style={font=\fontsize{6.5}{7.5}\selectfont, align=center},
]
\node[box, minimum width=16mm] (reg) at (0,0) {Registry\\\an{bridges, tokens,}\\[-1pt]\an{contracts per chain}};
\node[box, minimum width=20mm] (ana) at (29,0) {Stack analyzers\\\an{one per stack (7)}};
\node[box, minimum width=20mm] (rep) at (58,0) {Corridor report\\\an{tier per question,}\\[-1pt]\an{five caps}};
\node[box, minimum width=20mm] (tok) at (58,-19) {Token score\\\an{weakest answer}\\[-1pt]\an{per question}};
\node[box, minimum width=20mm] (mon) at (29,-19) {Monitor\\\an{watched changes}};
\draw[arr] (reg.east) -- node[above, lbl]{corridor} (ana.west);
\draw[arr] (ana.east) -- node[above, lbl]{reads at\\a pin} (rep.west);
\draw[arr] (rep.south) -- node[right, lbl]{every lane,\\$a{\to}b$, $b{\to}a$} (tok.north);
\draw[arr, dashed] (tok.west) -- node[below, lbl]{contracts} (mon.east);
\draw[arr] (mon.north) -- node[left, lbl]{rerun} (ana.south);
\end{tikzpicture}
\caption{Each corridor is scored by the analyzer of its bridge stack, and a
token takes the weakest answer on each question across its deployments and
lanes. Each stack's analyzer reads one pinned block per chain and reports a
tier per question and the five caps; the monitor reruns it when a watched
configuration changes.}
\label{fig:arch}
\end{figure}

\paragraph{\clarity{Architecture.}}
\clarity{We first collect every bridge and token on Monad, with their
contracts on every chain, into a \textit{registry} (Figure~\ref{fig:arch}). For each
bridge stack, we identify where its contracts store the data each rubric
question needs (Table~\ref{tab:evidence-map}). We turn each stack into its own
analyzer. Each corridor is assessed by the analyzer of the stack it uses. The
analyzer generates a report that scores the corridor by the rubric: a tier for
each question it can answer, and the five caps. All reads for one report come
from one fixed block per chain, a \emph{pin}, so a rerun at the same pins
gives the same report.}

\clarity{The analyzer answers the 17 questions that contract state and
on-chain events decide. The other nine scored questions depend on what happens
off chain: audits, public monitoring, the technology's past incidents, the
website and how the bridge's off-chain signers run. The four questions on
breaks by AI or quantum computers need reasoning about the cryptography behind
each privileged path, which fixed reads cannot do.}

\begin{table*}[tp]
\centering
\footnotesize
\setlength{\tabcolsep}{3pt}
\renewcommand{\arraystretch}{1.05}
\caption{\clarity{Each bridge stack stores its configuration differently in
its contracts, so the analyzer needs one adapter per stack. Each cell names
the function or setting the analyzer reads; ``--'' marks a setting it does not
read for that stack.}}
\label{tab:evidence-map}
\begin{tabularx}{\textwidth}{@{}l *{6}{>{\raggedright\arraybackslash}X}@{}}
\toprule
\textbf{Stack} & \textbf{Upgrade and admin} & \textbf{Verifiers and threshold} &
\textbf{Peers} & \textbf{Mint and backing} & \textbf{Rate limit} & \textbf{Pause} \\
\midrule
LayerZero V2 & app owner; endpoint delegate & required and optional DVNs and their count, from the receive library's config & peer per chain ID & simulated mint & app's limiter & app's pauser role, if any \\
LayerZero V1 & app owner & oracle and relayer, or DVNs & trusted remotes & simulated mint & app's caps & app's pauser role, if any \\
Chainlink CCIP & token pool owner & RMN signers and minimum; OffRamp $F$ of $n$ & supported chains, remote pools & minter roles; simulated mint & pool limiters & RMN curse \\
Wormhole NTT & manager owner & transceivers and threshold & peer per chain ID & manager mode: lock or burn & inbound and outbound limits & manager pauser \\
Wormhole Portal & proxy admin slot & guardian set, $\lfloor 2N/3\rfloor+1$ & registered bridges & wrapped-asset check & -- & -- \\
Circle CCTP v2 & owner; attester manager; token controller & enabled attesters and signature threshold & remote token messengers & minter allowance & burn limit per message & pauser \\
Hyperlane Warp & route owner & ISM tree: multisig validators and threshold, aggregation, routing & router per domain & collateral or synthetic route; mint limits & capacity and refill rate & pausable ISM \\
\bottomrule
\end{tabularx}
\end{table*}

\clarity{Some cases need explicit handling for correct scoring:}
\begin{itemize}[noitemsep,topsep=2pt,leftmargin=1.4em]
\item \clarity{\textit{Proxies.} Most bridge contracts are proxies: users call
  a fixed address that forwards to code an admin can replace. The owner of the
  proxy address can differ from that admin, so the analyzer reads the admin
  from the proxy's standard storage slots and scores upgrades by that admin's
  key.}
\item \clarity{\textit{EIP-7702 accounts.} EIP-7702 lets an ordinary account
  set code, so it looks like a contract, yet one key still controls it.
  Wherever a question asks whether a privileged account is a single key, the
  analyzer treats such an account as one.}
\item \clarity{\textit{Safes.} A Safe looks like a multisig, but a Safe that
  needs one signature lets any single signer act alone. The analyzer scores a
  Safe by how many signers must approve, so a one-signature Safe scores like a
  single key.}
\item \clarity{\textit{Timelocks.} A timelock delays upgrades so defenders can
  react, but some timelocks let a bypass role or an admin act at once.
  Counting only the delay would overstate the protection, so the analyzer
  scores each such role as a path with no delay.}
\item \clarity{\textit{LayerZero delegates.} A LayerZero app's owner can name a
  delegate that can change which verifiers approve the app's messages. Checking
  only the owner would miss this second way to swap the verifiers, so the
  analyzer scores verifier control by the weaker of the two.}
\item \clarity{\textit{Roles without a list.} Many contracts record who holds a
  role, such as the right to mint, but offer no way to list the holders. The
  analyzer rebuilds the list from the history of role grants up to the pin and
  checks that each holder still has the role, so a minter held by one key is
  found and scored.}
\end{itemize}

\paragraph{\clarity{Cross-chain analysis.}}
\label{sec:mesh}
\clarity{When a token's copies share one backing, a weak deployment on any
chain can empty it (Section~\ref{sec:rootcause}). A check from Monad misses a
weak link between two other chains, although supply minted through that link can then bridge into Monad.
Each such link, a \emph{lane}, has its own settings in each direction, so for
every pair of the token's chains $a$ and $b$ the analyzer checks both
$a{\to}b$ and $b{\to}a$, each at its own chain's pin. For $n$ chains this is
$n(n-1)$ lanes. The analyzer also follows each
deployment's configured peers and adds the deployments the registry missed.
The token's score takes, on each question, the weakest answer across all
deployments and lanes (Section~\ref{sec:rubric}).}

\paragraph{\clarity{Monitoring.}}
\label{sec:monitoring}
\clarity{An assessment holds only until the bridge's code or configuration
changes. Attackers make such changes on purpose: at Stake DAO the attacker
repointed the bridge's trusted peer minutes before the drain, and at DIMO the
drain came 24~seconds after a malicious upgrade
(Table~\ref{tab:monitoring-evidence}). We therefore extend the analyzer to
live monitoring. The monitor listens for
the events that change a deployment's authority, code, verifiers, supply or
limits. Each such event triggers a new analysis of the affected token at a
fresh pin. The alert's severity follows how the change moves the assessment:
a change that strengthens the bridge reads low, and one that weakens it reads
high. Where the bridge enforces a delay, a stop or a limit, an alert gives
defenders time to act before funds leave (Section~\ref{sec:damage}).}

\subsection{\clarity{Agentic Analyzer}}
\label{sec:agentic}
\clarity{We built our bespoke analyzers by hard-coding where each stack keeps
its contract information. This raises three issues: feasibility, security and
scalability. First, \textit{feasibility}: some checks need more than reading fixed
values. The questions on breaks by AI or quantum computers need a thorough
scan of the code behind each privileged path. Second, \textit{security}: a corridor may claim to use a known
stack yet change its contracts, for example by adding privileged roles through
other functions. Complex code can also hide a privileged path. Examples are a
diamond proxy split across many code modules, a call delegated to another
contract, a Safe module that acts without the owners' signatures and a role
granted without an event. A hard-coded analyzer reads only the places it
knows, so it can miss such a setting and report the corridor as safer than it
is. Third, \textit{scalability}: every contract change and every new bridge needs an analyzer
written or updated by hand.}

\clarity{An agentic analyzer avoids these issues by design: a language model
reads each bridge's code and state itself, so it can follow paths that no
fixed read anticipates, and it improves as language models improve. Our
agentic analyzer must therefore run the rubric's checks on any bridge,
driven by evidence, deterministically and at low cost. We state these four requirements
and then the architecture, with one part per requirement.}

\paragraph{\clarity{Requirements.}}
\clarity{The agentic analyzer must meet four requirements:}
\begin{itemize}[noitemsep,topsep=2pt,leftmargin=1.4em]
\item \clarity{\textit{Any bridge.} It runs the rubric's checks on a bridge it
  has not seen, with no code written for that bridge.}
\item \clarity{\textit{Evidence-driven.} It counts a fact only when recorded chain
  reads support it, since a language model can state a value it never read.}
\item \clarity{\textit{Determinism.} The same facts always give the same
  score, whatever model gathered them.}
\item \clarity{\textit{Low cost.} While the bridge's code and the data
  behind its answers are unchanged, it rechecks the bridge without the model,
  so a recheck spends no model tokens.}
\end{itemize}

\begin{figure}[tp]
\ifreview\color{CorrectionPurple}\captionsetup{font={color=CorrectionPurple}}\fi
\centering
\def\bbnm#1{{\fontsize{7.5}{8.5}\selectfont\bfseries #1}}
\begin{tikzpicture}[
  x=1mm, y=1mm,
  box/.style={draw=., text=., rounded corners=1pt, align=center,
    inner sep=1.3pt, fill=white, text width=32mm,
    font=\fontsize{7}{8}\selectfont},
  host/.style={box, fill=black!8},
  lbl/.style={text=., inner sep=.8pt, align=center,
    font=\fontsize{7}{8}\selectfont\itshape},
  ttl/.style={text=., inner sep=0pt,
    font=\fontsize{7.5}{8.5}\selectfont\bfseries},
  ar/.style={draw=., -{Latex[length=3pt,width=2.4pt]}, line width=.5pt},
  rp/.style={ar, dash pattern=on 2.2pt off 1.4pt},
  frame/.style={draw=., rounded corners=2pt}
]
\hyphenpenalty=10000 \exhyphenpenalty=10000 %
\node[host, text width=76mm] (rules) at (41.5,63)
  {\bbnm{Prompt}\\the rubric and the skills; the manifest as data};
\draw[frame, line width=1.1pt, fill=black!4] (0,32.5) rectangle (38,56.3);
\node[ttl, anchor=west] at (0,58.1) {Sealed container};
\node[box] (agent) at (19,51)
  {\bbnm{Agent (language model)}\\reasons, calls tools, submits typed facts};
\node[box] (disp) at (19,38.6)
  {\bbnm{Dispatcher}\\scope gate, reads, call IDs};
\node[host] (rec) at (19,27.0)
  {\bbnm{Call record}\\every call, recorded first};
\node[host] (proxy) at (19,17.0)
  {\bbnm{Egress proxy}\\the only route out};
\node[host] (chain) at (19,5.6)
  {\bbnm{Chains and explorers}\\pinned reads and source evidence};
\node[host] (anl) at (19,-5.0)
  {\bbnm{Analyzer reports}\\read log; same fact schema};
\node[host] (facts) at (64,51)
  {\bbnm{Typed facts}\\each value cites the calls behind it};
\node[host] (val) at (64,38.6)
  {\bbnm{Receipt validator}\\cited reads, required reads, model review};
\node[host] (score) at (64,25.6)
  {\bbnm{Host scorer}\\fixed rubric rules: tiers, caps};
\node[host] (gate) at (64,12.4)
  {\bbnm{Replay analyzers}\\written and reviewed once;\\same code hashes, fresh reads};
\draw[ar] ([xshift=-8.5mm]rules.south) -- node[lbl,right,pos=.5] {prompt}
  ([xshift=14mm]agent.north);
\draw[ar] (agent.south) -- (disp.north);
\draw[ar] ([xshift=9mm]disp.south) -- ([xshift=9mm]rec.north);
\draw[ar] (rec.south) -- (proxy.north);
\draw[ar] (proxy.south) -- (chain.north);
\draw[ar] (agent.east) -- node[lbl,above,align=center,xshift=2.2mm] {submit\\facts}
  (facts.west);
\draw[ar] (facts.south) -- node[lbl,right] {check} (val.north);
\draw[ar] (rec.east) -- (41.5,27.0) -- (41.5,41) -- (val.west |- 0,41);
\draw[ar] (val.south) -- node[lbl,right] {accepted facts} (score.north);
\draw[ar] (score.south) -- node[lbl,right] {capture} (gate.north);
\draw[rp] (gate.west) -- (43,12.4) -- (43,25) -- node[lbl,above] {replay}
  (rec.east |- 0,25);
\draw[ar] (anl.east) -- (45.5,-5.0) -- (45.5,36) -- (val.west |- 0,36);
\end{tikzpicture}
\caption{\clarity{The agentic analyzer has separate parts for each requirement.
The prompt's skills let the agent work on any bridge. The sealed container,
dispatcher, call record and receipt validator tie every fact to recorded
evidence. The host
scorer makes scores deterministic. Replay analyzers rerun without the model
at low cost (dashed path).}}
\label{fig:agentic-pipeline}
\end{figure}
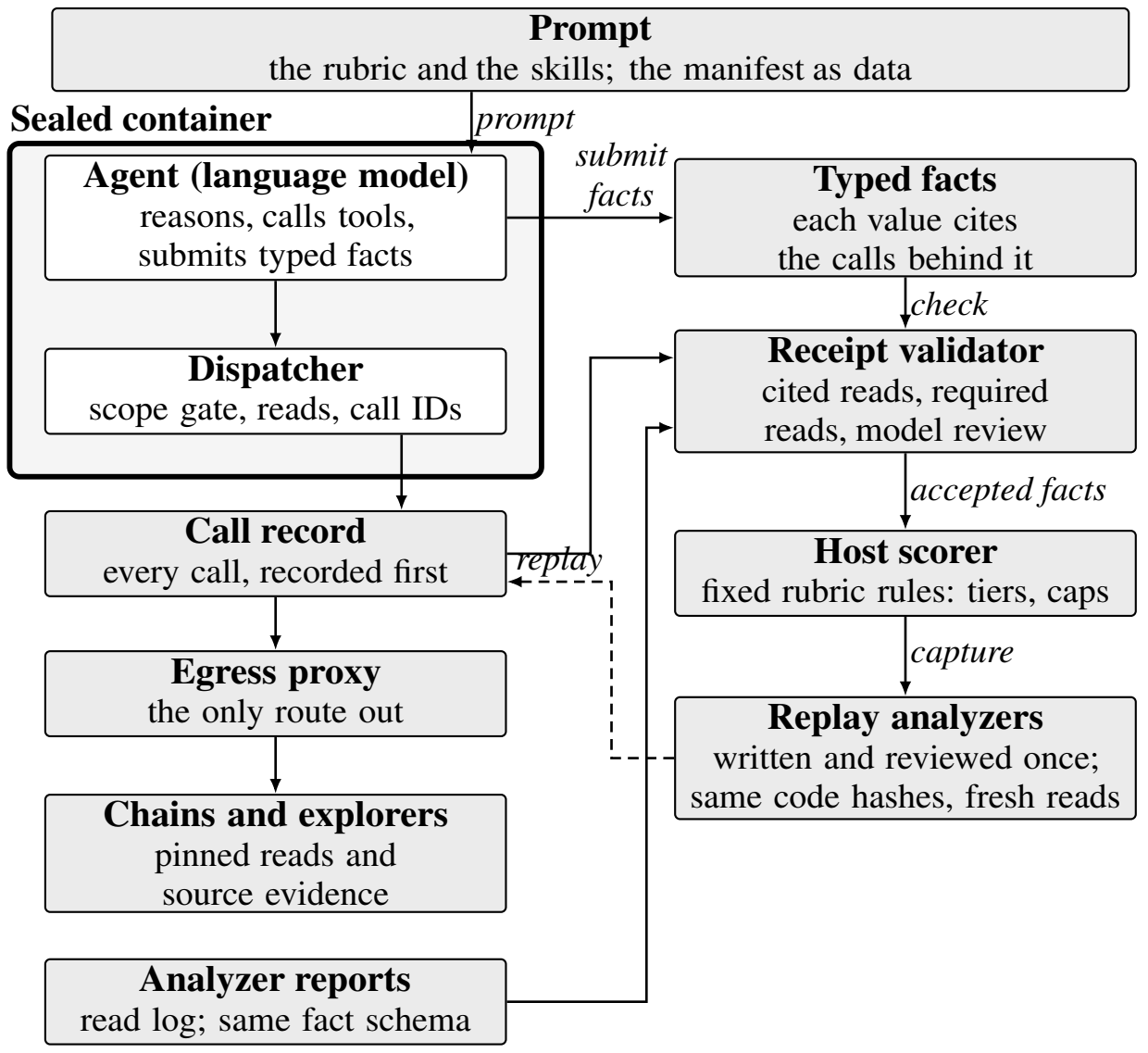

\paragraph{\clarity{Architecture.}}
\clarity{Figure~\ref{fig:agentic-pipeline} shows the architecture. We
describe it one requirement at a time.}

\clarity{\textit{Any bridge.} The agent gets no code written for the bridge.
Its only bridge-specific input is the bridge's \emph{manifest}: its contracts
on each chain, each labeled with what it is, such as the bridge contract, its
upgrade admin or the token. For the Monad tokens, we build the manifest from
the same registry as the bespoke analyzers. The agent's prompt holds the
\textit{rubric} and its \textit{skills}. A skill is a way to establish one kind
of fact:}
\begin{itemize}[noitemsep,topsep=2pt,leftmargin=1.4em]
\item \clarity{\textit{Resolve code}: follow a proxy to the code it runs, then
  read that code's verified source or decompile it.}
\item \clarity{\textit{Resolve authority}: find who can upgrade or reconfigure a
  contract, and tell a single key from a multisig with its threshold or a
  timelock with its delay.}
\item \clarity{\textit{Extract privileged roles}: find every function that
  changes code, settings or funds, the role allowed to call it and every
  account that holds that role, and confirm each holder from chain state.}
\item \clarity{\textit{Enumerate sets}: list every peer and verifier, from
  getters or the contract's event history.}
\item \clarity{\textit{Trace fund paths}: find each function that mints or
  releases funds, and the checks it applies.}
\item \clarity{\textit{Measure controls}: read rate limits, supply caps and
  pause state.}
\end{itemize}
\clarity{No skill, tool or label names a bridge stack; the skills find each
stack's functions from its contracts' code and events.}
\clarity{The agent has one tool, with which it either reads a contract at its
pin or submits a question's facts. An address that a read of an in-scope
contract returns joins the scope, so the agent can follow owners, Safes,
timelocks and role holders.}

\clarity{\textit{Evidence-driven.} A language model can report a value it never
read. Four controls keep such facts out. First, the agent runs in a
\textit{sealed container} with no credentials and no network access except a
proxy to the model provider. Every read therefore goes through the host's
\textit{dispatcher}, which records the call and returns a call ID. Second, each
fact must cite the IDs of the reads it depends on. The host's \textit{receipt
validator} checks the fact against those reads and rejects it if they do not
support it. Third, the host refuses a question whose facts lack the reads its
scoring rule needs. Fourth, a claim that a set is complete or that something
is absent also needs an independent model review of the cited reads.}

\clarity{\textit{Determinism.} The model never assigns a tier. The
\textit{host scorer} turns the accepted facts into tiers with fixed, ordered
rules, each quoting the rubric text it implements. It then folds the tiers
into deployment and token scores and applies the caps. The same facts
therefore always give the same score, whichever model gathered them. The
bespoke analyzers' facts enter the same scorer, so we can compare the two
analyzers.}

\clarity{\textit{Low cost.} Each investigation costs model tokens, so we turn
it into code that runs without the model. A \textit{writer model} turns each
question's accepted facts into a \emph{replay analyzer}, which recomputes them
from fresh reads, and a reviewer model must approve it. At a later pin, the
replay analyzer runs only if the code it depends on is unchanged. Otherwise
the question goes back to the agent (Section~\ref{sec:rq9}).}

\section{Evaluation}
\label{sec:eval}

\clarity{We proposed a security rubric for bridges and two kinds of analyzer
that apply it, bespoke and agentic. We first build a
counterfactual model that replays past hacks under defenses their bridges did
not have (Section~\ref{sec:eval-model}). With it, we evaluate whether the
rubric itself is correct: it must cover every past hack, its caps must fire on
the weaknesses that caused the hacks, and its tiers must capture real
differences in security. We also test whether containment really improves a
bridge's posture by reducing losses (Section~\ref{sec:cf-exp}). If these hold,
the next question is whether the rubric detects these weaknesses before the
attacks (Section~\ref{sec:preexploit}).}

\clarity{Finally, we evaluate the bespoke analyzers (Section~\ref{sec:rq5}) and
the agentic analyzer (Section~\ref{sec:rq7}).}

\subsection{\clarity{Counterfactual Model}}
\label{sec:eval-model}
\clarity{To evaluate the rubric against past hacks, we need a model of the
rubric's controls and containment levers together with the attacks. The model
must switch levers on and off against an attack and capture the loss each
time. With such a model, we evaluate counterfactually how well the rubric's
levers stop the attacks we have seen.}

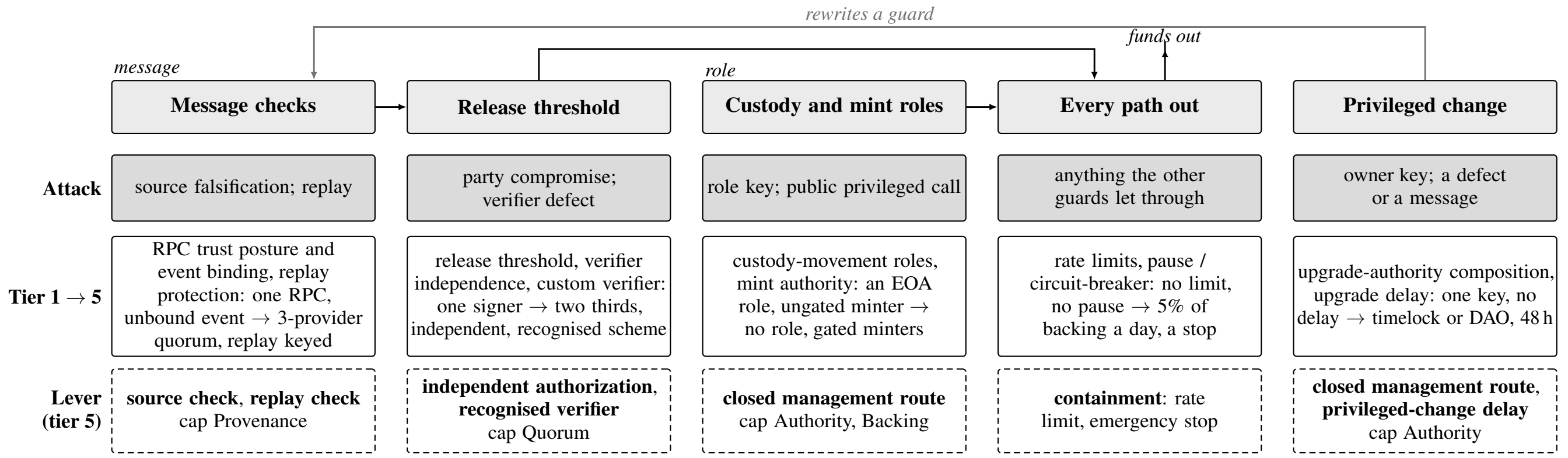
\begin{figure*}[tp]
\centering
\hyphenpenalty=10000 \exhyphenpenalty=10000
\begin{tikzpicture}[
  x=1mm, y=1mm,
  cell/.style={draw, rounded corners=1pt, align=center, inner sep=1.2pt,
    text width=29mm, minimum height=#1, font=\fontsize{6.4}{7.4}\selectfont},
  cell/.default=7mm,
  hd/.style={cell=6mm, fill=black!8, font=\fontsize{6.8}{7.8}\selectfont\bfseries},
  atk/.style={cell=7.5mm, fill=black!14},
  tier/.style={cell=13.6mm},
  lev/.style={cell=9.5mm, dash pattern=on 2pt off 1.2pt},
  rl/.style={inner sep=0pt, align=right, anchor=east, font=\fontsize{6.6}{7.6}\selectfont\bfseries},
  ar/.style={-{Latex[length=2.6pt,width=2.2pt]}, line width=.55pt},
  fl/.style={inner sep=.8pt, font=\fontsize{6.3}{7.3}\selectfont\itshape},
]
\node[hd] (h1) at (27,42) {Message checks};
\node[hd] (h2) at (60.5,42) {Release threshold};
\node[hd] (h3) at (94,42) {Custody and mint roles};
\node[hd] (h4) at (127.5,42) {Every path out};
\node[hd] (h5) at (161,42) {Privileged change};
\draw[ar] (h1) -- (h2);
\draw[ar] (h2.north) |- ++(0,3.5) -| node[fl, above, pos=.25] {} ([xshift=-4mm]h4.north);
\draw[ar] (h3) -- (h4);
\node[fl, anchor=south west] at (h1.north west) {message};
\node[fl, anchor=south west] at (h3.north west) {role};
\draw[ar] ([xshift=4mm]h4.north) |- ++(0,3.5) node[fl, anchor=south] {funds out};
\draw[ar, black!55] (h5.north) |- ++(0,6) -| node[fl, above, pos=.25, text=black!55] {rewrites a guard} ([xshift=8mm]h1.north);
\node[rl] at (11,32.8) {Attack};
\node[atk] at (27,32.8) {source falsification; replay};
\node[atk] at (60.5,32.8) {party compromise; verifier defect};
\node[atk] at (94,32.8) {role key; public privileged call};
\node[atk] at (127.5,32.8) {anything the other guards let through};
\node[atk] at (161,32.8) {owner key; a defect or a message};
\node[rl] at (11,20.5) {Tier 1 $\to$ 5};
\node[tier] at (27,20.5) {RPC trust posture and event binding, replay protection: one RPC, unbound event $\to$ 3-provider quorum, replay keyed};
\node[tier] at (60.5,20.5) {release threshold, verifier independence, custom verifier: one signer $\to$ two thirds, independent, recognised scheme};
\node[tier] at (94,20.5) {custody-movement roles, mint authority: an EOA role, ungated minter $\to$ no role, gated minters};
\node[tier] at (127.5,20.5) {rate limits, pause / circuit-breaker: no limit, no pause $\to$ 5\% of backing a day, a stop};
\node[tier] at (161,20.5) {upgrade-authority composition, upgrade delay: one key, no delay $\to$ timelock or DAO, 48\,h};
\node[rl] at (11,7.5) {Lever\\(tier 5)};
\node[lev] at (27,7.5) {\textbf{source check}, \textbf{replay check}\\cap Provenance};
\node[lev] at (60.5,7.5) {\textbf{independent authorization}, \textbf{recognised verifier}\\cap Quorum};
\node[lev] at (94,7.5) {\textbf{closed management route}\\cap Authority, Backing};
\node[lev] at (127.5,7.5) {\textbf{containment}: rate limit, emergency stop};
\node[lev] at (161,7.5) {\textbf{closed management route}, \textbf{privileged-change delay}\\cap Authority};
\end{tikzpicture}
\caption{\clarity{Each guard of the modeled bridge has an attack that meets it,
rubric questions whose tiers set it, and a lever that its top tier switches on.
A message passes message checks and the release threshold, a role skips both,
every path out passes containment, and a privileged change can rewrite any
guard. A hijacked website passes no guard.}}
\label{fig:cf-model}
\end{figure*}

\clarity{To build it, we encode three things: each hacked bridge, its attack
and the levers. The bridge's elements come from the rubric. The attack's
capabilities are new, because the rubric scores bridges and has no attacker.
Each bridge comes with its own attack, and each run switches the levers on or
off. The model
builds on timed attack-defense trees~\cite{kordy2011foundations,gadyatskaya2016timed}: each path's requirement is an attack tree
with threshold gates, and levers act as countermeasures fixed before the
attack. Unlike those trees, time is a fixed list of events and paths share
funds and an allowance, so we compute loss on runs. Figure~\ref{fig:cf-model} shows all three parts together.}

\paragraph{\clarity{Bridge elements.}}
\clarity{We encode each bridge, as its postmortem documents it, by the elements
the rubric scores. A configuration $C=(P,K,G,\mathit{Act},\Pi,V)$ holds:}
\begin{itemize}[noitemsep,topsep=1pt,leftmargin=1.2em]
\item \clarity{\textit{Parties} $P$ and \textit{keys} $K$ (People \& Power):
  each key $k$ has one owner $o(k)\in P$, so a set $Q\subseteq P$ of compromised
  parties exposes every key whose owner is in $Q$.}
\item \clarity{\textit{Threshold groups} $G$ (People \& Power for authority,
  Code \& Config for release): a group
  $g=(K_g,m_g)$ holds under $Q$ when at least $m_g$ keys of $K_g$ have owners
  in $Q$.}
\item \clarity{\textit{Privileged actions} $\mathit{Act}$ (Code \& Config): an
  action $a=(\varphi_a,d_a)$, such as an upgrade or a new verifier set, changes
  code or settings once its requirement $\varphi_a$ holds and its delay $d_a$
  has passed.}
\item \clarity{\textit{Transfer paths} $\Pi$ (Code \& Config): a path
  $\pi=(\varphi_\pi,v_\pi,b_\pi,r_\pi,\sigma_\pi,\rho_\pi)$ releases or mints value from
  reserve $v_\pi$ when $\varphi_\pi$ holds, spending rate limit $b_\pi$ (none
  for a path outside the bridge's contracts, $a_\bot=\infty$) on route $r_\pi$. The requirement $\varphi_\pi$
  combines groups and actions with AND and OR; $\sigma_\pi$ marks a source
  check and $\rho_\pi$ a replay check. A \textit{message path} pays when a
  cross-chain message passes the verifiers, a \textit{role path} when a
  privileged role moves custody or mints with no message, and a
  \textit{pool path} when a swap pays out of a bridge pool.}
\item \clarity{\textit{Reserves} $V$ (Backing): a reserve $v$ holds the value
  $X_v$ a path can take, such as an escrow's balance for a release path or the
  supply a minter may still create for a mint path. Some hacks take users' own
  funds instead: a bridge contract spends their standing approvals, or a
  hijacked website sends their payments to the attacker.}
\end{itemize}

\paragraph{\clarity{Attack capabilities.}}
\clarity{The rubric has no attacker, so we define an attacker $A=(Q,f)$. Its
parties $Q\subseteq P$ hold their keys, and $f\in\{0,1\}$ is 1 when its message
passes a source check: what it credits really happened on the source
chain, or a defect writes the state the check reads. The catalogue's
mechanism families form four classes of capability:}
\begin{itemize}[noitemsep,topsep=1pt,leftmargin=1.2em]
\item \clarity{\textit{Party compromise}: stolen or extracted keys of verifiers
  or of a role, or a shared signer control plane; it adds parties to $Q$.}
\item \clarity{\textit{Privileged change}: an upgrade, or a new verifier set,
  peer, owner or minter. A stolen owner key runs it through the owner's delay;
  a defect or a cross-chain message that reaches the change skips the owner
  and its delay.}
\item \clarity{\textit{Code defect}: a verifier or proof bypass, a public
  privileged call, an accounting flaw, an arbitrary call that spends users'
  approvals, or a replay of a used authorization. It enters $Q$ as a one-key
  party that meets the one requirement where the defect sits.}
\item \clarity{\textit{Source falsification}: a poisoned source view fools the
  verifiers that read it. An event a contract really emitted, or a field the
  bridge never binds to the source, fools every honest verifier, so only a
  source check rejects it.}
\end{itemize}
\clarity{Each hack is encoded once from its postmortem: its paths, whether a
privileged change came first, and how the attacker met each requirement. No
encoding says which lever stops the hack.}

\paragraph{\clarity{Levers.}}
\clarity{Each lever is the top tier of the rubric questions in parentheses, and
each run switches levers on or off. The first five are \emph{preventive
checks}:}
\begin{itemize}[noitemsep,topsep=1pt,leftmargin=1.2em]
\item \clarity{\textit{Independent authorization} (release threshold, verifier
  independence): every message and role path also needs an uncompromised,
  independent key, or the threshold $m_g$ rises to its tier-5 value; a
  message every honest verifier would sign still passes, but a poisoned RPC
  view does not, since an independent verifier reads its own.}
\item \clarity{\textit{Closed management route} (upgrade-authority
  composition, verifier-set control, custody-movement roles, mint authority, code
  mutability): no privileged change runs, and no role moves custody or mints.}
\item \clarity{\textit{Source check} (RPC trust posture and event binding,
  per-chain finality, reorg / attestation-cancel): sets $\sigma_\pi$, which the attacker cannot pass with $f=0$.}
\item \clarity{\textit{Replay check} (replay protection): sets $\rho_\pi$.}
\item \clarity{\textit{Recognised verifier} (custom verifier): a verifier
  defect no longer meets the release threshold.}
\item \clarity{\textit{Rate limit} $b=(c_b,r_b,p_b)$ (rate limits, destination supply
  cap):
  an allowance of at most $c_b$ that refills by $r_b$ every period $p_b$,
  shared by every path through which value leaves the bridge's contracts.}
\item \clarity{\textit{Emergency stop} (pause / circuit-breaker, bunker
  mode): halts every
  path through the bridge's contracts and cancels pending privileged actions.}
\item \clarity{\textit{Privileged-change delay} (upgrade delay): sets
  $d_a=48$\,h on changes made with the owner's key, so a change waits before it takes effect and a stop can cancel
  it.}
\end{itemize}
\clarity{\emph{Containment} is the rate limit, the emergency stop and the
privileged-change delay of Section~\ref{sec:damage}.}

\paragraph{\clarity{Tiers and caps.}}
\clarity{Each tier of Section~\ref{sec:rubric} fixes one setting of a guard,
and each cap marks a setting so weak that one compromised party or one false
message reaches the funds. Figure~\ref{fig:cf-model} places each setting and
cap on its guard.}

\paragraph{\clarity{Transitions.}}
\clarity{An attack unfolds as a timed list of events $E=(e_1,\ldots,e_N)$ acting
on a state $s=(t,X,a,D,H,U,\ell)$: the time, the reserve balances, the allowances, the
times at which privileged actions were authorized, the stopped routes, the
used message identifiers and the outflow so far. Each event changes $s$ as
follows:}
\begin{itemize}[noitemsep,topsep=1pt,leftmargin=1.2em]
\item \clarity{\textit{Time}: before an event at time $t'$, each allowance
  refills to $a_b\leftarrow\min\{c_b,\ a_b+r_b(\lfloor t'/p_b\rfloor-\lfloor
  t/p_b\rfloor)\}$, and $t\leftarrow t'$.}
\item \clarity{$\mathit{authorize}(\alpha)$: if $\varphi_\alpha$ holds under
  $Q$ and no stop blocks $\alpha$, then $D(\alpha)\leftarrow t$. The action
  takes effect once $t\ge D(\alpha)+d_\alpha$.}
\item \clarity{$\mathit{transfer}(\pi,q,id)$ releases or mints $q$ on path
  $\pi$ only if its guard $\Gamma_\pi$ holds, and then updates $s$:}
\end{itemize}
\[
\begin{aligned}
\Gamma_\pi={}&(Q\models\varphi_\pi)\land r_\pi\notin H\land(\sigma_\pi\Rightarrow f)\\
  &\land(\rho_\pi\Rightarrow id\notin U)\land q\le\min\{X_{v_\pi},a_{b_\pi}\},\\
s'={}&s\,[\,X_{v_\pi}\!-\!q,\ a_{b_\pi}\!-\!q,\ U\cup\{id\},\ \ell+q\,].
\end{aligned}
\]
\begin{itemize}[noitemsep,topsep=0pt,leftmargin=1.2em]
\item \clarity{$\mathit{stop}(r)$ adds route $r$, or every route through the bridge's contracts, to $H$; a stop of every route also cancels pending authorizations; $\mathit{resume}(r)$ removes $r$ from $H$ but
  restores no cancelled authorization.}
\end{itemize}
\clarity{Here $Q\models\varphi_\pi$ means the requirement holds with the keys
of $Q$ and the actions that have taken effect. A rejected event changes
nothing. The \emph{outflow} $L(C,A,E)$ is $\ell$ after the last event.
A run starts at $t=0$ with full reserves and allowances, no authorizations
or stopped routes, $\ell=0$, and $U$ holding the identifiers used before the
attack. In each hack's $E$, the attacker authorizes its change at $t=0$, then every path in turn asks for the full allowance at the
start, for the refill at the stop time before the stop, for the full allowance just after it, and, if it needs the change, after the 48\,h delay. Reserves are set above every request.}

\clarity{To compute the outflow, we compile $C$, $A$ and $E$ into the input
language of the PRISM model checker~\cite{kwiatkowska2011prism}. PRISM
explores every state the events reach, with exact whole-number amounts, so
the outflow comes from an exhaustive search instead of sampled runs.}

\paragraph{\clarity{Validation.}}
\label{sec:model-validation}
\clarity{PRISM gives the hand-calculated outflow in 41 small cases and matches
a second implementation on 1,560 synthetic inputs and every hack. That
implementation matches small Solidity contracts on a local EVM chain in eight
transfers.}

\subsection{\clarity{Counterfactual Experiments}}
\label{sec:cf-exp}
\clarity{We run two kinds of experiment on the model: hack replays test
coverage and caps and measure loss cuts, and tier comparisons test whether
higher tiers lose less.}

\paragraph{\clarity{Hack replays.}}
\clarity{We replay every encoded hack 11 times in PRISM~4.11 with its explicit
engine: with no lever, with containment, with its rate limit and stop alone,
with the five preventive checks together and each alone, with \emph{full
controls} (every lever on), and with the levers of the questions its postmortem
names. A hack that fires a cap runs a 12th time, with the levers of the
questions behind its caps. Each run records the outflow and, for each request, whether it passed
and which guard blocked it.}

\clarity{For our experiments, containment uses a \$\IdealControlAllowanceM{}M
allowance that refills \$\IdealControlRefillM{}M a day and a stop after
\IdealControlStopHours{} hour, a limit stricter than tier~5 requires of a
bridge holding more than \$40M. The losses we compute therefore rest on this
assumption.}

\paragraph{\clarity{Coverage.}}
\label{sec:rq2}\label{sec:evaluation}
\clarity{A hack is covered when the top tier of the question its postmortem
names stops it in the model. Of the \CoverageTotalN{} hacks, that tier stops
\CoverageNamedStopsN{}. For \CoverageNamedMissN{}, it misses and containment
caps the loss: Thorchain's attacker held the real vault key, SagaEVM's forged
deposits came through the chain's own precompile, and Pike's misbound release
needed a source check. For \CoverageUnnamedContainedN{}, the postmortem names no
question and containment caps the loss. The last two are Celer, whose hijacked
website sent users' payments straight to attacker contracts, which no lever
meters, and one hack that only threatened the bridge's availability.}

\paragraph{\clarity{Caps.}}
\label{sec:caps}
\clarity{Caps fire on \CorpusCapsN{} of the \CorpusPostureN{} posture failures:
Quorum on \CorpusQuorumN{}, Authority on \CorpusAuthorityN{}, Backing on
\CorpusBackingN{} and Provenance on \CorpusProvenanceN{}, and several failures
fire more than one cap. PRISM confirms that the caps mark the setting that let
the funds out: with only the questions behind its caps at their top tier,
every one of the \CapCheckCappedN{} hacks loses nothing. The remaining
failure, Celer's hijacked website, gets the lowest tier of the website
question, which has no cap.}

\paragraph{\clarity{Tiers.}}
\label{sec:validation}
\clarity{Of the rubric's 84 tier steps, \PhysicalTierBranches{} change what an
attacker can take. On each, PRISM runs one setting of the lower and one of the higher tier against the same declared set of attacks, and the higher tier never loses more and
loses less in at least one attack; this does not cover every setting a tier
allows. The other
\the\numexpr\EvidenceTierBoundaries+\UnestablishedTierBoundaries\relax{} leave
losses unchanged, and each is justified by what it gives users instead. Of these, \EvidenceTierBoundaries{} give
public evidence: 13 about the code that runs, 7 about incidents as they
happen, 9 about the bridge's record, 3 about its source chain's rules and 2
about its backing. The last \UnestablishedTierBoundaries{} keep the bridge
available.}

\paragraph{\clarity{Losses.}}
\label{sec:rq4}\label{sec:counterfactual}
\clarity{We set aside \IdealControlZeroN{} code bugs that lost nothing:
researchers reported five before any attack, and validators stopped the sixth
before it paid out. We replay the other \LossHacksN{} hacks, which lost
\$\CorpusGrossB{}B. The rubric's controls target posture
(Section~\ref{sec:rootcause}), so the \PostureFailuresN{} posture failures face
full controls and the \LossCodeN{} code hacks face containment alone. Losses
fall \IdealControlReductionFold{}-fold, to \$\IdealControlResidualM{}M:}
\begin{itemize}[noitemsep,topsep=1pt,leftmargin=1.2em]
\item \clarity{\textit{Stopped}: \IdealControlPreventedN{} posture failures. The
  control behind each failure, at its top tier, blocks every path the attacker
  used.}
\item \clarity{\textit{Limited by containment}: \IdealControlBoundedN{} code
  hacks, each capped at the smaller of \$1.04M and what it took,
  \$\IdealControlBoundedM{}M in total.}
\item \clarity{\textit{Unaffected}: Celer, \$\IdealControlOutsideM{}M. Its hijacked
  website sent users' payments straight to attacker contracts, so no bridge
  contract, and no lever, ever touched the funds.}
\end{itemize}
\clarity{Containment alone on every hack cuts losses \ContainmentCreditFold{}-fold, to
\$\ContainmentCreditResidualM{}M\label{sec:financial-account}. With a stop
after a day instead of an hour, the cut is still
\IdealControlDayReductionFold{}-fold.}

\subsection{Detection from on-chain data before the exploit}
\label{sec:preexploit}

\begin{clarityblock}
Coverage asks whether the rubric names each hack's weakness, using the
postmortem. Here we ask whether on-chain data alone shows that weakness before
the attack. Code bugs are not part of this test, since a posture check does
not look for bugs (Section~\ref{sec:rootcause}). For each of the
\CorpusPostureN{} posture failures, we read only the bridge's contracts, at a
pin before the exploit, and check whether the cap its postmortem names fires.

The cap fires for \CorpusPreExploitCapsN{} of the \CorpusPostureN{}. These
lost \$\CorpusPreExploitCapsM{}M of the \$\CorpusPostureM{}M that posture
failures lost. The other \the\numexpr\CorpusPostureN-\CorpusPreExploitCapsN\relax{}
had their weakness off chain, so on-chain data cannot show it. Five had relay
software that accepted fake events: pNetwork~\cite{pm-pnetwork-pbtc}, THORChain
twice~\cite{pm-thorchain-bifrost,pm-thorchain-router}, TAC~\cite{pm-tac} and
Alephium~\cite{pm-alephium}. Two had one operator holding the signing keys in
one place: Mixin~\cite{pm-mixin} and AFX~\cite{pm-afx-primary}. The last is
Celer's frontend~\cite{pm-celer}.
\end{clarityblock}

\subsection{Analyzer correctness on live bridges}
\label{sec:rq5}
\label{sec:applied}
\label{sec:rq6}

\begin{clarityblock}
We ran the analyzer on the \EvalBoardCorridors{} tokens and
shared each report with the bridge technology provider and the team that runs
the deployment. We fixed every analyzer defect they confirmed. The most common
objection was that a key held by an externally owned account is protected off
chain, by a multisig or MPC. Chain state shows what an address may do, not how
its key is kept, so we keep the finding and mark it cautionary.

\paragraph{Cross-chain assessment.}
We ran the cross-chain analyzer on the 37 tokens on 1 October 2026, excluding
non-EVM chains. It assessed 190 deployments on 16 EVM chains, 153 of them off
Monad, and 798 lanes between them, 603 of which never touch Monad. It answered
2,707 of the 2,821 applicable questions on these deployments (96\%).
\end{clarityblock}

\subsection{Agentic analyzer}
\label{sec:rq7}\label{sec:rq8}

\begin{clarityblock}
We test the agentic analyzer against its four requirements
(Section~\ref{sec:agentic}). Determinism: on the same bridges, it must give the
bespoke analyzer's answers. Evidence-driven: its evidence checks must reject
false claims. Any bridge: it must find the needed facts on bridges it has not
seen. Low cost: it must repeat an unchanged assessment without the model. We
also run it on the AI-threat and quantum questions, which the bespoke
analyzers cannot answer.

\paragraph{Determinism: agreement with the analyzer.}
On all \EvalBoardCorridors{} Monad corridors, the agent and the bespoke
analyzer each answer every question. Of the \EvalBoardAnalyzerDecided{} answers
the analyzer decides, the agent gives the same answer on
\EvalBoardDecisionMatches{} (\EvalBoardAgentAgreementPct{}\%,
Table~\ref{tab:agent-agreement}); an answer the agent leaves unresolved counts
as a miss.
\end{clarityblock}

\begin{table}[tp]
\centering\footnotesize
\caption{\clarity{The agent gives the bespoke analyzer's answer on
\EvalBoardAgentAgreementPct{}\% of the answers the analyzer decides. Each cell
counts answers where the agent gave the row's tier and the analyzer the
column's; the diagonal holds the \EvalBoardTierMatches{} matches.}}
\label{tab:agent-agreement}
\begin{tabular}{r|rrrrr}
 & \multicolumn{5}{c}{Bespoke analyzer's tier}\\
Agent's tier &1&2&3&4&5\\\hline
1 & 176 & 0 & 0 & 0 & 1\\
2 & 0 & 119 & 0 & 0 & 3\\
3 & 0 & 1 & 106 & 0 & 3\\
4 & 1 & 0 & 1 & 4 & 0\\
5 & 1 & 0 & 3 & 0 & 96\\
\end{tabular}

\end{table}

\begin{clarityblock}
\paragraph{Evidence-driven: evidence checks.}
We take facts the host accepted in nine completed agent sessions. For each,
we build a false copy that changes one thing, such as the contract the fact is
about, the block it was read at, its value or a call it cites, and keep the
original as a control. We send both to the session's checker and to the host's
receipt check. Across 183 such pairs, covering ten kinds of change, both checks
reject every false copy and accept every original. A probe of the container
also confirms that the session cannot read the host's files, write to
read-only folders, reach the network directly or call the scorer.

\paragraph{Any bridge: discovery on unfamiliar bridges.}
We give the agent only the rubric, a hacked bridge's contract addresses and a
pin before the exploit: no stack name and no labels. None of these bridges uses
one of the seven stacks the agent was built on. On the
\HistoricalFilteredAgentN{} hacks we ran, it named the weakness the attacker
used in every one: four missing rate limits and four weak authority, release or
mint settings. Without the per-stack read routes and stack names, and with one
generic skill instead, all answers on two corridors were correct.

\paragraph{AI threat and quantum resistance.}
\label{sec:pq-results}
These questions ask whether a key broken by AI-found mathematics or a quantum
computer could still complete a privileged action. The bespoke analyzers
cannot answer them. On 12 cases across all seven stacks, the agent found five
gaps an operator can fix, such as a pause its EOA owner can lift.

\paragraph{Low cost: no model call when the bridge is unchanged.}
\label{sec:rq9}
We test whether replay analyzers repeat an assessment without the model. On
nine corridors, we rerun them 20.9 days later at new pins: each checks that the
code it depends on is unchanged, re-reads the state and lets the host score.
All 135 question assessments replay: 133 with no model call and the same
facts and tiers, and two go back to the agent because the bridge changed in
between, as designed.
\end{clarityblock}

\section{Discussion}
\label{sec:discussion}
\begin{clarityblock}
Our results point to three things operators can do now.

\paragraph{Fix the caps first.}
A cap marks the setting the funds left through. In our model, raising only
the questions behind each hack's caps to their top tier stops all
\CapCheckCappedN{} capped hacks (Section~\ref{sec:caps}). Most of these
settings were visible before the attack: on-chain data alone shows the cap for
\CorpusPreExploitCapsN{} of the \CorpusPostureN{} posture failures, which lost
\$\CorpusPreExploitCapsM{}M of \$\CorpusPostureM{}M
(Section~\ref{sec:preexploit}).

\paragraph{Delay every privileged change, and limit every path out.}
Attackers who take a key often change the bridge and drain it within the hour,
as in \CorpusTimingWithinHourN{} of the \CorpusTimingN{} hacks with known
timing (Figure~\ref{fig:drain}). An alert alone comes too late. A 48-hour
delay with a stop cancels the change before it takes effect in
\ContainmentCreditPreventedN{} of our replays. The rate limit must cover every
path out of the bridge's contracts, not only its transfers and mints: 9 of the
\IdealControlBoundedN{} code hacks that containment caps were routers or
gateways spending users' standing approvals.

\paragraph{Check what chains cannot show.}
All \the\numexpr\CorpusPostureN-\CorpusPreExploitCapsN\relax{} posture
failures that on-chain data missed sat off chain: five in relay software, two
in signing keys that one operator held in one place, and one in a frontend.
Operators should publish evidence others can check for these parts, such as
how their signing keys are held. Teams most often told us that off-chain MPC
protects their single-key accounts, which chain data cannot confirm.
\end{clarityblock}

\section{Limitations and Threats to Validity}
\label{sec:limits}
\begin{clarityblock}
We state what our assessment cannot see, then what could make our results
wrong inside and outside our data.

\paragraph{Limitations.}
We read on-chain data. Relay software, key storage and frontends sit off
chain, so for them we give operators a checklist; this is why on-chain data
missed \the\numexpr\CorpusPostureN-\CorpusPreExploitCapsN\relax{} of the
\CorpusPostureN{} posture failures (Section~\ref{sec:preexploit}). Chain data
also cannot show how a single-key account's key is held, so we mark such
findings cautionary. The scored questions target bridges that trust a committee of verifiers;
light-client, optimistic and solver bridges need other caps, which this
version does not score. Our live reads cover EVM chains only, and issuer powers
such as freezing a token stay out of scope. Both are engineering scope, not
limits of the method: a reader for another chain or for issuer roles plugs into
the same pipeline, and the cross-chain findings already cover every EVM chain a
token reaches.

\paragraph{Internal validity.}
We built the caps from the hacks we test them on, so the caps show that the
rubric expresses known failures, not that it predicts new ones. Labeling a hack
posture or code takes judgment. Our main finding, that posture failures cause
most losses since 2023, still holds if we drop KelpDAO
(\CorpusWithoutKelpPct{}\%) or count off-chain software failures as code, as
surveys do (\CorpusBoundaryPct{}\%)~\cite{lee2023sok,notland2024sok,wu2025bridgeguard}.

The loss results come from a model, so each depends on how we encoded the
hack. Two reviewers who read only the postmortems agree with the model's
outcome on 20 of 22 hacks we checked. The cuts also depend on the containment
setting we chose, a \$1M allowance and a one-hour stop: at \$5M and a one-day
stop the cut falls to \IdealControlMinFold{}-fold. We use reported losses
before recoveries; after recoveries, posture failures still hold
\CorpusAdjustedPct{}\% of the loss.

We measure the agent against the bespoke analyzer, so the agent's accuracy
is only as good as the analyzer's: where both are wrong, the test cannot see
it. We also built the agent on those same bridges. A replay repeats its
original investigation, errors included.

\paragraph{External validity.}
Our catalogue holds publicly documented hacks, so it leans toward large,
reported losses, and 2026 covers January to July. Without bridges that were
never hacked~\cite{liu2015cloudy,zhang2014mismanagement}, we cannot show that a
low score predicts a hack. The loss cuts assume each control works as modeled:
a rate limit that one key can lift, or that misses a path out, saves less.
Future attacks may use failures our past hacks do not show. Bridge scores rest on tiers tested against losses, but they still carry our
judgment, most where no loss test applies: the
\the\numexpr\EvidenceTierBoundaries+\UnestablishedTierBoundaries\relax{} tier
steps that leave losses unchanged, and the category weights of 40, 35, 25 and
15 that combine the questions into one score. Others can set the weights to
whatever they judge better.
\end{clarityblock}

\section{Conclusion}
\label{sec:conclusion}
\begin{clarityblock}
Bridge hacks have cost billions, and since 2023
\CorpusRecentPct{}\% of reported losses came from posture failures,
not code bugs. We turned these hacks into a rubric that scored deployed
bridges from on-chain evidence, marked fatal settings with hard caps and
followed each token across every deployment that shared its backing. A
counterfactual model replayed every hack: the rubric's controls stopped
\IdealControlPreventedN{} of the \PostureFailuresN{} posture failures, a rate
limit and an emergency stop capped the code hacks, and together they cut losses
\IdealControlReductionFold{}-fold, from \$\CorpusGrossB{}B to
\$\IdealControlResidualM{}M. Our analyzer applied the rubric to the
\EvalBoardCorridors{} tokens listed on Monad and watched them for changes. Our
agentic analyzer handled bridges it had not seen, tolerated hallucination by
matching every fact to recorded evidence, and let the model write each replay
program once and then retire, reusing the program until the bridge changed.

Future work can include a way for operators to prove how their off-chain
relay software and keys are run, a fully proved formalism of the bridge model,
and automatic lifting of bridge smart contracts into the model, so any live
bridge can be replayed under the levers.
\end{clarityblock}

\label{endofbody}

\appendix
\section{Incident Catalogue}
\label{app:catalogue}
Table~\ref{tab:corpus} lists the \CorpusN{} incidents used in the analysis and
\CorpusAdjacentN{} adjacent incidents, for a catalogue of \CorpusCatalogueN{}.
The adjacent cases, Telcoin, Velocore, Solv and Radiant, are excluded from the
totals because their recorded losses did not require a failed cross-chain
authorization. \clarity{Shibarium's 2023 launch halt is excluded because no one
attacked it: the chain stalled under load and deposits waited until it
recovered.} The supplementary catalogue records each incident's cause,
supporting evidence, countermeasures and primary sources. It distinguishes
transaction links to an established exploit from links to a privileged
configuration change.

\onecolumn
\footnotesize
\begin{longtable}{@{}p{2.5cm}lp{1.9cm}rll@{}}
\caption{Complete incident corpus, most recent first. \correction{Reported gross under the catalogue conventions;} ``wh''=whitehat near-miss. \correction{Source confidence from the supplementary catalogue: M, medium (11); MH, medium-high (2), among the \CorpusN{} in-scope entries. Six labelled posture incidents carry \$29.2M reported gross.} \correction{The transaction column carries the exploit transaction where a source establishes one and, where none does, the privileged configuration change the entry turns on; the role is named in the supplementary catalogue. Hashes are truncated and, where the chain is determinable, link to the transaction on its block explorer. ``n/a'' marks a row for which no consulted source publishes a transaction in either role, including near-misses.} \correction{Layer 1 (L1) and layer 2 (L2) identify blockchain layers; Syscoin uses the Network-Enhanced Virtual Machine (NEVM).}}\label{tab:corpus}\\
\toprule
\textbf{Bridge} & \textbf{Date} & \textbf{Chains} & \correction{\textbf{Reported gross}} & \textbf{Class} & \correction{\textbf{Transaction}} \\ \midrule
\endfirsthead
\toprule \textbf{Bridge} & \textbf{Date} & \textbf{Chains} & \correction{\textbf{Reported gross}} & \textbf{Class} & \correction{\textbf{Transaction}} \\ \midrule
\endhead
\midrule \multicolumn{6}{r}{\emph{continued on next page}}\\ \endfoot
\bottomrule \endlastfoot
Verus-Ethereum Bridge\textsuperscript{M} & 2026-07-23 & Verus, Ethereum & \$7.5M & Verification bug & \correction{n/a} \\
AFX Trade bridge\textsuperscript{M} & 2026-07-22 & Arbitrum, Ethereum & \$24M & Quorum compromise & \correction{n/a} \\
Wanchain (NIGHT)\textsuperscript{M} & 2026-07-20 & Cardano, BNB Chain & \$9.0M & Verification bug & \correction{n/a} \\
Allbridge Core\textsuperscript{M} & 2026-07-19 & Solana, Ethereum & \$1.65M & Oracle manipulation & \correction{n/a} \\
Aztec Connect Bridge\textsuperscript{M} & 2026-06-14 & Ethereum, Aztec & \$4.2M & Verification bug & \correction{n/a} \\
Syscoin Bridge\textsuperscript{M} & 2026-06-07 & Syscoin, Syscoin NEVM & \$8.6M & Verification bug & \correction{n/a} \\
Gravity Bridge & 2026-05-30 & Ethereum, Cosmos & \$5.4M & Quorum compromise & \correction{n/a} \\
Alephium TokenBridge & 2026-05-30 & Ethereum, BNB Chain +1 & \$815k & Attester forgery & \correction{n/a} \\
Stake DAO vsdCRV bridge & 2026-05-27 & Ethereum, Arbitrum & \$91k & Admin-key compromise & \correction{\href{https://arbiscan.io/tx/0x7489ec5f5dba1de6e6c92f2c0f1dd93bd4a2f307c3bd2305b2f93f569a3e5fe5}{\texttt{0x7489ec5f}\ldots}} \\
Squid\-Router\-Module & 2026-05-25 & Ethereum, Base +1 & \$4.0M & Verification bug & \href{https://etherscan.io/tx/0x59d17fd31e31959b2d562508bf91c4fc1271682ba7d61a6209865e1151b69aea}{\texttt{0x59d17fd3}\ldots} \\
Butter Bridge V3.1\textsuperscript{MH} & 2026-05-21 & MAP Relay Chain, Ethereum +1 & \$180k & Verification bug & \href{https://etherscan.io/tx/0x31e56b4737649e0acdb0ebb4eca44d16aeca25f60c022cbde85f092bde27664a}{\texttt{0x31e56b47}\ldots} \\
Verus-Ethereum Bridge & 2026-05-18 & Verus, Ethereum & \$12M & Verification bug & \href{https://etherscan.io/tx/0xb428dae60a234c149c8bc4468979356c434726bf8f588b6c97bd7144aa5bcefe}{\texttt{0xb428dae6}\ldots} \\
THORChain & 2026-05-15 & THORChain, Bitcoin +9 & \$11M & Admin-key compromise & \href{https://etherscan.io/tx/0x4370739cf3f443fe129727ea1a9e215783d881c643f3ea1d12ce822aeb3e6af8}{\texttt{0x4370739c}\ldots} \\
Adshares Bridge & 2026-05-15 & Adshares, Ethereum & \$628k & Attester forgery & \href{https://etherscan.io/tx/0xa3476575183204b4a662dd6ee56f6499d806e4f41ce83d98366752d31e9e9ca3}{\texttt{0xa3476575}\ldots} \\
TAC\textsuperscript{M} & 2026-05-11 & TON, TAC +1 & \$2.9M & Attester forgery & \correction{n/a} \\
ZetaChain Gateway & 2026-04-26 & ZetaChain, Ethereum +3 & \$334k & Access-control bug & \href{https://etherscan.io/tx/0x81fc9b2457b3ea66e2cefe2afe65d083ce358a11520e0f4aa5faad0bfea18a56}{\texttt{0x81fc9b24}\ldots} \\
KelpDAO rsETH OFT bridge & 2026-04-18 & Ethereum, Unichain +2 & \$292M & Attester forgery & \href{https://etherscan.io/tx/0x1ae232da212c45f35c1525f851e4c41d529bf18af862d9ce9fd40bf709db4222}{\texttt{0x1ae232da}\ldots} \\
Hyperbridge & 2026-04-13 & Ethereum, Base +3 & \$2.5M & Verification bug & \href{https://etherscan.io/tx/0x240aeb9a8b2aabf64ed8e1e480d3e7be140cf530dc1e5606cb16671029401109}{\texttt{0x240aeb9a}\ldots} \\
Aethir ATH OFT bridge\textsuperscript{MH} & 2026-04-09 & BNB Chain, Tron +1 & \$90k & Access-control bug & \correction{n/a} \\
ioTube & 2026-02-21 & Ethereum, IoTeX +3 & \$4.4M & Admin-key compromise & \href{https://etherscan.io/tx/0xe9e7f33ebfe2230c147e6e0321f5f2c7de1b89fe9fc08830fc3f8ac5845bc9f0}{\texttt{0xe9e7f33e}\ldots} \\
CrossCurve & 2026-02-01 & Ethereum, Arbitrum +8 & \$3.0M & Verification bug & \href{https://etherscan.io/tx/0x37d9b911ef710be851a2e08e1cfc61c2544db0f208faeade29ee98cc7506ccc2}{\texttt{0x37d9b911}\ldots} \\
SagaEVM IBC precompile bridge & 2026-01-21 & SagaEVM, Ethereum mainnet & \$7.0M & Verification bug & \href{https://sagaevm.sagaexplorer.io/tx/0x0c038d70c684b5797ed5b8ac578cf7151ec95f5a1a135cd9d48028f72d0f7a2b}{\texttt{0x0c038d70}\ldots} \\
DIMO Network cross-chain bridge\textsuperscript{M} & 2025-11-06 & Ethereum, Polygon +2 & \$40k & Admin-key compromise & \href{https://etherscan.io/tx/0xaa22f3b50ea8a4bbacc4e58e5bc5b5cae3ba1c2a50a216f8d1c724ec052bd082}{\texttt{0xaa22f3b5}\ldots} \\
Garden Finance & 2025-10-30 & Bitcoin, Ethereum +5 & \$11M & Quorum compromise & \correction{n/a} \\
402bridge\textsuperscript{M} & 2025-10-27 & Base, Arbitrum & \$18k & Admin-key compromise & \href{https://basescan.org/tx/0x089a6336425c6ee6d8954923763cbaeef1173ce44b5c0ab853c85863726e46e2}{\texttt{0x089a6336}\ldots} \\
Seedify SFUND cross-chain bridge\textsuperscript{M} & 2025-09-23 & Avalanche, Ethereum +3 & \$1.7M & Admin-key compromise & \correction{n/a} \\
Yala \$YU LayerZero OFT & 2025-09-13 & Polygon, Solana +1 & \$7.6M & Admin-key compromise & \correction{n/a} \\
Shibarium Bridge & 2025-09-12 & Shibarium, Ethereum & \$2.4M & Quorum compromise & \href{https://etherscan.io/tx/0xe882a83afb92d6070b848ef025ae699ec043b7c2f31b21d2a08c94306f9b817e}{\texttt{0xe882a83a}\ldots} \\
Force Bridge & 2025-06-01 & Ethereum, BNB Chain +1 & \$3.8M & Quorum compromise & \href{https://etherscan.io/tx/0x6b6fbd9d6beef56d2a4f0d14852beea381764b962d7d73ecd216b9fd991299a1}{\texttt{0x6b6fbd9d}\ldots} \\
FEG SmartBridge & 2024-12-29 & Ethereum, BNB Smart Chain +1 & \$1.0M & Verification bug & \href{https://bscscan.com/tx/0x6c3cf48e9c8b22f60fc2b702b03f93926ee8624639b100fe851b757947e61759}{\texttt{0x6c3cf48e}\ldots} \\
Ronin Bridge V2 & 2024-08-06 & Ethereum, Ronin & \$12M & Init/upgrade bug & \href{https://etherscan.io/tx/0x2619570088683e6cc3a38d93c3d98899e5783864e15525d5f5810c11189ba6cb}{\texttt{0x26195700}\ldots} \\
LI.FI & 2024-07-16 & Ethereum, Arbitrum & \$9.7M & Router drain & \correction{\href{https://etherscan.io/tx/0xd82fe84e63b1aa52e1ce540582ee0895ba4a71ec5e7a632a3faa1aff3e763873}{\texttt{0xd82fe84e}\ldots}} \\
Holograph Protocol & 2024-06-13 & Ethereum, Mantle & \$14M & Access-control bug & \href{https://etherscan.io/tx/0x0cc143ccf3316d47b36a2e45577922f4ebe2374966bb22c1e9cf49c747d46396}{\texttt{0x0cc143cc}\ldots} \\
XLink & 2024-05-14 & BNB Smart Chain, Ethereum +2 & \$4.3M & Admin-key compromise & \href{https://bscscan.com/tx/0x47e123af93add709bc2516f6a5db057dfbb1d66a75b693cd7980cd3eb28c7357}{\texttt{0x47e123af}\ldots} \\
Pike Finance & 2024-04-30 & Ethereum, Arbitrum +1 & \$2.0M & Init/upgrade bug & \correction{n/a} \\
XBridge & 2024-04-24 & Ethereum, BNB Chain & \$1.4M & Access-control bug & \href{https://etherscan.io/tx/0x903d88a92cbc0165a7f662305ac1bff97430dbcccaa0fe71e101e18aa9109c92}{\texttt{0x903d88a9}\ldots} \\
Merkle Trade Omnichain Swap \& \ldots & 2024-04-20 & Ethereum, Arbitrum +5 & \$19k & Router drain & \href{https://arbiscan.io/tx/0x462cab29471473852a7743cfd5cf3c0cdbaa7e05a85a3431bfca3668c50fa4dd}{\texttt{0x462cab29}\ldots} \\
Socket Protocol & 2024-01-16 & Ethereum & \$3.3M & Router drain & \href{https://etherscan.io/tx/0xc6c3331fa8c2d30e1ef208424c08c039a89e510df2fb6ae31e5aa40722e28fd6}{\texttt{0xc6c3331f}\ldots} \\
Orbit Bridge & 2023-12-31 & Ethereum, Orbit Chain & \$82M & Quorum compromise & \correction{\href{https://etherscan.io/tx/0x958aeec58ea2f0f9700adda24e43fb76f9e052e4c20773f180c49d7529d95f16}{\texttt{0x958aeec5}\ldots}} \\
HECO Bridge & 2023-11-22 & Ethereum, HECO +2 & \$87M & Admin-key compromise & \href{https://etherscan.io/tx/0xbb6fe88427c2f3bc179075109d47a805dcfedab0e475eaca0d979311873e131b}{\texttt{0xbb6fe884}\ldots} \\
Mixin Network & 2023-09-23 & Ethereum, Bitcoin +1 & \$200M & Quorum compromise & \correction{n/a} \\
zkSync Era canonical L1/L2 b\ldots & 2023-09-19 & Ethereum, zkSync Era & wh/\$0 & Verification bug & \correction{n/a} \\
Shibarium Bridge & 2023-08-16 & Ethereum, Shibarium & wh/\$0 & Liveness halt & \correction{n/a} \\
Multichain & 2023-07-06 & Fantom, Ethereum +3 & \$126M & Quorum compromise & \href{https://etherscan.io/tx/0x448f2a6a6c071cdce254937e06305a033538e1aeb9339227d0e59e0458e6185c}{\texttt{0x448f2a6a}\ldots} \\
Poly Network & 2023-07-01 & Ethereum, BNB Chain +9 & \$4.4M & Quorum compromise & \href{https://etherscan.io/tx/0xc0361cd88cb734774db760f3aac2cd7477f59a0221a2c0ff8266318b7d6317a8}{\texttt{0xc0361cd8}\ldots} \\
Allbridge Core & 2023-04-01 & BNB Chain & \$574k & Oracle manipulation & \href{https://bscscan.com/tx/0x7ff1364c3b3b296b411965339ed956da5d17058f3164425ce800d64f1aef8210}{\texttt{0x7ff1364c}\ldots} \\
Rubic & 2022-12-25 & Ethereum, Arbitrum +5 & \$1.4M & Router drain & \href{https://etherscan.io/tx/0x9a97d85642f956ad7a6b852cf7bed6f9669e2c2815f3279855acf7f1328e7d46}{\texttt{0x9a97d856}\ldots} \\
pNetwork & 2022-11-03 & Ethereum, BNB Chain / BSC & \$4.6M & Admin-key compromise & \correction{n/a} \\
QANX Bridge & 2022-10-11 & BNB Smart Chain, Ethereum & \$1.2M & Quorum compromise & \href{https://bscscan.com/tx/0xf93047e41433d73ddf983cfa008aeb356ec89803c0a92b0e97ccdc6c42a13f51}{\texttt{0xf93047e4}\ldots} \\
BSC Token Hub & 2022-10-06 & BNB Beacon Chain, BNB Smart Chain +6 & \$569M & Verification bug & \href{https://bscscan.com/tx/0xebf83628ba893d35b496121fb8201666b8e09f3cbadf0e269162baa72efe3b8b}{\texttt{0xebf83628}\ldots} \\
Transit Swap & 2022-10-01 & BNB Chain, Ethereum +1 & \$21M & Router drain & \href{https://bscscan.com/tx/0x181a7882aac0eab1036eedba25bc95a16e10f61b5df2e99d240a16c334b9b189}{\texttt{0x181a7882}\ldots} \\
Arbitrum One canonical bridge & 2022-09-19 & Ethereum, Arbitrum One & wh/\$0 & Init/upgrade bug & \correction{n/a} \\
Celer cBridge & 2022-08-17 & Ethereum, BSC +8 & \$235k & Frontend hijack & \href{https://etherscan.io/tx/0x246c6e14891182cf0dfda3a2e509874306b464680f9853f72c12194c4c5e8451}{\texttt{0x246c6e14}\ldots} \\
Nomad & 2022-08-01 & Ethereum, Moonbeam +4 & \$190M & Init/upgrade bug & \correction{\href{https://etherscan.io/tx/0xa5fe9d044e4f3e5aa5bc4c0709333cd2190cba0f4e7f16bcf73f49f83e4a5460}{\texttt{0xa5fe9d04}\ldots}} \\
Harmony Horizon Bridge & 2022-06-23 & Harmony, Ethereum +1 & \$100M & Quorum compromise & \href{https://etherscan.io/tx/0x27981c7289c372e601c9475e5b5466310be18ed10b59d1ac840145f6e7804c97}{\texttt{0x27981c72}\ldots} \\
EvoDeFi Bridge & 2022-06-07 & Oasis Emerald, BNB Smart Chain +10 & \$66M & Admin-key compromise & \href{https://bscscan.com/tx/0xba5c0add385bf81c91201e0a6d1cf5c7ea794ce3e52eae69445cc00461f22285}{\texttt{0xba5c0add}\ldots} \\
Rainbow Bridge / Aurora Engine & 2022-04-26 & Aurora, NEAR +1 & wh/\$0 & Verification bug & \correction{n/a} \\
Ronin Bridge & 2022-03-23 & Ethereum, Ronin & \$540M & Quorum compromise & \href{https://etherscan.io/tx/0xc28fad5e8d5e0ce6a2eaf67b6687be5d58113e16be590824d6cfa1a94467d0b7}{\texttt{0xc28fad5e}\ldots} \\
LI.FI & 2022-03-20 & Ethereum & \$600k & Router drain & \href{https://etherscan.io/tx/0x4b4143cbe7f5475029cf23d6dcbb56856366d91794426f2e33819b9b1aac4e96}{\texttt{0x4b4143cb}\ldots} \\
Wormhole & 2022-02-24 & Ethereum & wh/\$0 & Init/upgrade bug & \correction{n/a} \\
Meter Passport & 2022-02-05 & Ethereum, BNB Smart Chain +2 & \$4.3M & Verification bug & \href{https://bscscan.com/tx/0xc7eb98e00d21ec2025fd97b8a84af141325531c0b54aacc37633514f2fd8ffdc}{\texttt{0xc7eb98e0}\ldots} \\
Wormhole & 2022-02-02 & Solana, Ethereum & \$320M & Verification bug & \correction{n/a} \\
Optimism canonical bridge & 2022-02-02 & Ethereum, Optimism +2 & wh/\$0 & Verification bug & \correction{n/a} \\
Qubit Finance QBridge & 2022-01-27 & Ethereum, BNB Smart Chain & \$80M & Verification bug & \href{https://etherscan.io/tx/0x478d83f2ad909c64a9a3d807b3d8399bb67a997f9721fc5580ae2c51fab92acf}{\texttt{0x478d83f2}\ldots} \\
Multichain & 2022-01-18 & Ethereum, BNB Chain +3 & \$3.0M & Router drain & \href{https://etherscan.io/tx/0xe50ed602bd916fc304d53c4fed236698b71691a95774ff0aeeb74b699c6227f7}{\texttt{0xe50ed602}\ldots} \\
Synapse Protocol & 2021-11-06 & Avalanche, Polygon +2 & wh/\$0 & Oracle manipulation & \correction{n/a} \\
Polygon & 2021-10-05 & Ethereum, Polygon PoS & wh/\$0 & Verification bug & \correction{n/a} \\
pNetwork & 2021-09-19 & BNB Chain, Bitcoin & \$13M & Attester forgery & \correction{\href{https://bscscan.com/tx/0x0eb55e02bce39ec1d2d2e911eca7dcca54e74841b53412c078185e43c5a2a551}{\texttt{0x0eb55e02}\ldots}} \\
Poly Network & 2021-08-10 & Ethereum, Binance Smart Chain +1 & \$611M & Access-control bug & \href{https://etherscan.io/tx/0xb1f70464bd95b774c6ce60fc706eb5f9e35cb5f06e6cfe7c17dcda46ffd59581}{\texttt{0xb1f70464}\ldots} \\
THORChain & 2021-07-23 & Ethereum, THORChain & \$7.9M & Attester forgery & \href{https://etherscan.io/tx/0x2e9611f4b3a4729005a00653197b7f26211a9d8d20c1bdf193613b58dc6af93b}{\texttt{0x2e9611f4}\ldots} \\
ChainSwap & 2021-07-11 & Ethereum, BSC +2 & \$4.4M & Access-control bug & \href{https://etherscan.io/tx/0xb582e7d1dc8d456d52f6b2b394942fd1406b1bf02ce97e909efbc699caf41974}{\texttt{0xb582e7d1}\ldots} \\
Anyswap & 2021-07-10 & Ethereum, BNB Chain +1 & \$7.9M & Quorum compromise & \href{https://etherscan.io/tx/0xecaaf8b57b6587412242fdc040bd6cc084077a07f4def24b4adae6fbe8254ae3}{\texttt{0xecaaf8b5}\ldots} \\
THORChain\textsuperscript{M} & 2021-06-29 & Ethereum, THORChain & \$350k & Attester forgery & \correction{n/a} \\
\midrule \multicolumn{6}{@{}l}{\textit{Adjacent / out-of-scope (cross-chain context, excluded from totals)}}\\ \midrule
Solv Protocol & 2026-03-05 & Ethereum & \$2.7M & Verification bug & \href{https://etherscan.io/tx/0x44e637c7d85190d376a52d89ca75f2d208089bb02b7c4708ad2aaae3a97a958d}{\texttt{0x44e637c7}\ldots} \\
Radiant Capital & 2024-10-16 & Arbitrum, BSC +2 & \correction{\$50M} & Admin-key compromise & \correction{\href{https://arbiscan.io/tx/0x7856552db409fe51e17339ab1e0e1ce9c85d68bf0f4de4c110fc4e372ea02fb1}{\texttt{0x7856552d}\ldots}} \\
Velocore & 2024-06-02 & zkSync Era, Linea +1 & \$6.8M & Access-control bug & \href{https://lineascan.build/tx/0xed11d5b013bf3296b1507da38b7bcb97845dd037d33d3d1b0c5e763889cdbed1}{\texttt{0xed11d5b0}\ldots} \\
Telcoin & 2023-12-25 & Polygon PoS & \$1.2M & Init/upgrade bug & \href{https://polygonscan.com/tx/0x35f50851c3b754b4565dc3e69af8f9bdb6555edecc84cf0badf8c1e8141d902d}{\texttt{0x35f50851}\ldots} \\
\end{longtable}
\normalsize
\twocolumn

Table~\ref{tab:monitoring-evidence} documents the selected privileged changes
and subsequent drains used in Figure~\ref{fig:drain}.
\begin{table*}[tp]
\centering
\footnotesize
\setlength{\tabcolsep}{4pt}
\renewcommand{\arraystretch}{1.0}
\begin{corrected}
\caption{Corpus incidents whose loss a privileged on-chain configuration change
 enabled, ordered by the interval between the change and the drain. These are
 detection opportunities, not demonstrations that a deployed monitor prevented a
 drain. Each interval carries its evidence: explorer-verified transactions,
 single-source block numbers, a reported window, or a date-level documentary record.
 Truncated hashes identify the transactions recorded in the supplementary catalogue;
 ``n/a'' marks a transaction not selected in the retained record.}
\label{tab:monitoring-evidence}
\begin{tabularx}{\textwidth}{@{}l X l l p{3.1cm} r@{}}
\toprule
\textbf{Incident} & \textbf{Trackable change (on-chain event)} & \textbf{Change tx} &
\textbf{Drain tx} & \textbf{Change\,$\to$\,drain (basis)} & \textbf{Reported gross} \\
\midrule
DIMO (2025-11-06) & \texttt{upgrade} to malicious implementation, then
 \texttt{emergencyWithdraw}~\cite{pm-dimo} & \texttt{0xaa22f3b5}\ldots &
 \texttt{0x274edfde}\ldots & 24\,s (explorer-verified, blocks 23{,}742{,}033 and
 23{,}742{,}035) & \$40k \\
Stake DAO (2026-05-27) & \texttt{setPeer} repoints the OFT to an attacker
 contract~\cite{pm-stakedao} & \texttt{0x59fd50ce}\ldots & \texttt{0x7489ec5f}\ldots & ${\sim}25$\,s from the last re-set, 227\,s
 from the first (retained RPC logs) & \$91k \\
402bridge (2025-10-27) & \texttt{transferOwnership} of the settlement contract, then
 \texttt{transferUserToken} against outstanding approvals~\cite{pm-402bridge} &
 \texttt{0x089a6336}\ldots & n/a & minutes (reported; sweep over about 26 min) &
 \$18k \\
ioTube (2026-02-21) & \texttt{transferOwnership}, then \texttt{upgrade} transfers
 vault and mint ownership~\cite{pm-iotube} & \texttt{0xe9e7f33e}\ldots,
 \texttt{0xc9c53b28}\ldots & \texttt{0x1b5772b2}\ldots & 13 min or more (takeover transactions 13 min apart;
 drain later) & \$4.4M \\
XLink (2024-05-14) & four to five malicious implementation upgrades of the bridge
 endpoint proxy~\cite{pm-xlink} & n/a & n/a & under 1 h (reported) & \$4.3M \\
Yala (2025-09-13) & leftover deployment key \texttt{setPeer}s a counterfeit peer;
 dormant~\cite{pm-yala} & n/a & n/a & 32 d (documentary, date-level) & \$7.6M \\
EvoDeFi (2022-06-07) & \texttt{setOperatorMode} grants three \texttt{taker}
 roles that send escrow out directly~\cite{pm-evodefi} & \texttt{0xba5c0add}\ldots & n/a &
 164--166 d (documentary, date-level: roles set 2021-12-23) & \$66M \\
\bottomrule
\end{tabularx}
\end{corrected}
\end{table*}

\section{Retrospective Evidence}
\label{app:retrospective-evidence}
Tables~\ref{tab:eval_per_incident} and~\ref{tab:eval_per_incident_b} list,
for each of the \CorpusPostureN{} posture failures, the rubric questions and
caps that its postmortem names. Section~\ref{sec:preexploit} reads each of
these caps from chain state at a pin before the exploit: \CorpusPreExploitCapsN{}
fire, and the other \the\numexpr\CorpusPostureN-\CorpusPreExploitCapsN\relax{}
sit off chain. The evidence labels mark whether the attribution rests on chain
observations (C), documents (P) or both (C/P).

\begin{table*}[tp]
\centering\footnotesize
\setlength{\tabcolsep}{4pt}
\renewcommand{\arraystretch}{1.0}
\caption{Every posture incident draws at least one question, and \CorpusCapsN{} of the \CorpusPostureN{}
draw a cap. Part 1 of 2: admin-key and verification-quorum incidents, grouped by
failure mechanism. Rows retain the retrospective
record's order within each group. Group headings describe failure mechanisms
without restricting question membership: Garden's Q6 row is retained among the
privileged-key cases. In QANX, \texttt{signers[dep]} = max means the deployer's
stored signing limit is $2^{256}-1$, an unlimited signing limit. \emph{Basis} is \textbf{C}, an
on-chain observation reported at the cited pre-exploit block; \textbf{P},
documentary evidence; \textbf{C/P}, both. A basis of \textbf{C} records the
observation and not a complete replay of the cap predicate: six of the \CorpusBasisReadN{}
\textbf{C} and \textbf{C/P} rows include a complete pinned command;
five report a block without a complete command
(Appendix~\ref{app:retrospective-evidence}).
Cloud-service abbreviations: Amazon Web Services (AWS), Key Management Service (KMS).}
\label{tab:eval_per_incident}
\begin{tabularx}{\textwidth}{@{}>{\raggedright\arraybackslash}p{1.85cm} l X >{\raggedright\arraybackslash}p{1.55cm} c >{\raggedright\arraybackslash}p{4.6cm}@{}}
\toprule
\textbf{Bridge} & \textbf{Date} & \textbf{Failure and the question that flags it} & \textbf{Cap} & \textbf{Basis} & \textbf{Evidence} \\
\midrule
\multicolumn{6}{@{}l}{\textbf{Admin / privileged-key compromise: owner, timelock, residual roles (Q1, Q2, Q4)}}\\
pNetwork GALA & 2022-11-03 & pGALA \texttt{ProxyAdmin} owner EOA, unlimited mint (Q1, Q8) & Authority, Backing & \textbf{C} & \href{https://bscscan.com/address/0xF8C69b3A5DB2E5384a0332325F5931cD5Aa4aAdA}{\texttt{0xF8C6\ldots}}\,@\,\allowbreak 22745013~\allowbreak ProxyAdmin \texttt{owner}\,$=$\,\allowbreak EOA \cite{pm-pgala} \\
HECO Bridge & 2023-11-22 & one un-rotated operator EOA on withdraw path, no timelock (Q1, Q2, Q6) & Authority, Quorum & \textbf{P} & two operator keys, one un-rotated holder \cite{pm-heco} \\
XLink (ALEX Lab) & 2024-05-14 & Bridge Endpoint \texttt{ProxyAdmin} one EOA, no multisig or timelock (Q1) & Authority & \textbf{C} & \href{https://bscscan.com/address/0xb3955302E58FFFdf2da247E999Cd9755f652b13b}{\texttt{0xb395\ldots}}\,@\,\allowbreak 38720023~\allowbreak ProxyAdmin \texttt{owner}\,$=$\,\allowbreak EOA \cite{pm-xlink} \\
Seedify SFUND & 2025-09-23 & OFT \texttt{owner} a dev EOA; mint ungated by inbound message (Q1, Q8) & Authority, Backing & \textbf{P} & stolen dev key minted ungated SFUND \cite{pm-seedify} \\
402bridge & 2025-10-27 & bridge \texttt{owner} a deployer EOA, key on a server (Q1) & Authority & \textbf{C} & \href{https://basescan.org/address/0xed1AFc4DCfb39b9ab9d67f3f7f7d02803cEA9FC5}{\texttt{0xed1A\ldots}}\,@\,\allowbreak 37396597~\allowbreak \texttt{owner()}\,$=$\,\allowbreak EOA \cite{pm-402bridge} \\
Garden Finance & 2025-10-30 & solver fills on one hot key, no co-signer or threshold (Q6) & Quorum & \textbf{P} & single solver hot key, no co-signer \cite{pm-garden} \\
DIMO Network & 2025-11-06 & NTT-manager \texttt{owner} one EOA (Q1) & Authority & \textbf{C} & \href{https://etherscan.io/address/0x07C64bd1B23b7C9B0ABf80b8613f58e5B00ED5dD}{\texttt{0x07C6\ldots}}\,@\,\allowbreak 23742032~\allowbreak \texttt{owner()}\,$=$\,\allowbreak EOA \cite{pm-dimo} \\
ioTube & 2026-02-21 & one EOA owns upgradeable validator: release, mint, verifiers (Q1, Q3, Q8) & Authority, Quorum, Backing & \textbf{C} & \href{https://etherscan.io/address/0xE7eBA1CEA51EC9B3AcCC16728e3B8786560c59d5}{\texttt{0xE7eB\ldots}}\,@\,\allowbreak 24501846~\allowbreak \texttt{owner()}\,$=$\,\allowbreak EOA \cite{pm-iotube} \\
Stake DAO vsdCRV & 2026-05-27 & vsdCRV OFT \texttt{owner} (slot 0) one deployer EOA, \texttt{setPeer}s trusted peer (Q1, Q3) & Authority, Quorum & \textbf{C} & \href{https://arbiscan.io/address/0x62d5a59E0d67c0381aAd53B201B4A1B8Dcd2C833}{\texttt{0x62d5\ldots}}\,@\,\allowbreak 467160930~\allowbreak \texttt{owner}\,$=$\,\allowbreak EOA \cite{pm-stakedao} \\
Yala \$YU OFT & 2025-09-13 & leftover deploy key wires counterfeit peer, mints unbacked \$YU, no cap (Q3, Q9) & Backing, Quorum & \textbf{P} & leftover deploy key wired rogue peer to mint \cite{pm-yala} \\
\addlinespace[2pt]
\multicolumn{6}{@{}l}{\textbf{Verification-quorum compromise: threshold and independence (Q6, Q7)}}\\
Anyswap Router V3 & 2021-07-10 & MPC signing collapses to one non-diverse key, recovered by nonce reuse (Q6, Q7) & Quorum & \textbf{P} & MPC key fully recovered (nonce reuse) \cite{pm-anyswap} \\
Ronin Bridge & 2022-03-23 & 5-of-9 met; four Sky Mavis keys plus un-revoked delegation, $N$ not independent (Q7) & Quorum & \textbf{P} & Sky Mavis controlled 5 of 9 signers \cite{pm-ronin} \\
Harmony Horizon & 2022-06-23 & sub-majority release multisig, \texttt{required()}\,$=$\,2 of 4 (Q6) & Quorum & \textbf{C} & \href{https://etherscan.io/address/0x715CdDa5e9Ad30A0cEd14940F9997EE611496De6}{\texttt{0x715C\ldots}}\,@\,\allowbreak 15012645~\allowbreak \texttt{required()}\,$=$\,\allowbreak 2/4 \cite{pm-harmony} \\
Poly Network II & 2023-07-01 & whole keeper quorum exfiltrated from one build env (Q7) & Quorum & \textbf{P} & keeper keys exfiltrated from one build env \cite{pm-polynetwork2} \\
Multichain & 2023-07-06 & all MPC nodes in one operator cloud account (Q7) & Quorum & \textbf{P} & all MPC nodes on one personal cloud account \cite{pm-multichain} \\
Mixin Network & 2023-09-23 & reported concentration of signing material in one cloud database (Q6, Q7) & Quorum & \textbf{P} & cloud compromise reported; exact signing boundary uncertain \cite{pm-mixin} \\
Orbit Bridge & 2023-12-31 & reported signer concentration (Q7); compromise mechanism disputed & Quorum & \textbf{P} & valid release signatures; operator disputes validator-key theft \cite{pm-orbit} \\
Force Bridge & 2025-06-01 & 3-of-4 validator multisig, 3 sigs on-chain, all Magickbase-run; keys stolen off-chain (Q7) & Quorum & \correction{\textbf{C/P}} & \correction{\href{https://etherscan.io/address/0x63A993502e74828ddba5710327AFC6dc78d661b2}{\texttt{0x63A9\ldots}}\,@\,\allowbreak 22608305~\allowbreak \texttt{multisigThreshold\_()}\,$=$\,\allowbreak 3} \cite{pm-forcebridge} \\
Shibarium (2025) & 2025-09-12 & 10-of-12 validator keys in one AWS KMS account; stake flash-loanable (Q7) & Quorum & \textbf{P} & 10 of 12 keys in one AWS KMS account \cite{pm-shibarium2025} \\
Gravity Bridge & 2026-05-30 & on-chain 2/3 voting power passed; keys stolen off-chain, mechanism undocumented (Q7) & \correction{Unresolved} & \textbf{P} & 2/3 of signer keys stolen; independence undisclosed \cite{pm-gravity} \\
THORChain (TSS) & 2026-05-15 & \correction{nominal} $t$-of-$n$ TSS; churned-in node rebuilds key via \texttt{tss-lib} GG20 flaw (Q6, Q7) & \correction{None assigned} & \textbf{P} & vault key rebuilt via \texttt{tss-lib} GG20 flaw \cite{pm-thorchain-tss} \\
AFX Trade & 2026-07-22 & legitimate 2/3 threshold over seven validators AFX ran itself, one internal Ansible control plane privileged on all, so independence fails (Q6, Q7) & Quorum & \textbf{P} & 5 of 7 validator keys compromised through that control plane; quorum passed \cite{pm-afx-primary} \\
QANX Bridge & 2022-10-11 & deployer key seeds \texttt{signers}; one EOA owns release (Q1, Q6) & Authority, Quorum & \textbf{C} & \href{https://etherscan.io/address/0xAAA4940aA878D932D3482Bf1DE332E1D50c15AaA}{\texttt{0xAAA4\ldots}}\,@\,\allowbreak 15723621~\allowbreak \texttt{signers[dep]}\,$=$\,\allowbreak max \cite{pm-qanx} \\
\bottomrule
\end{tabularx}
\end{table*}

\begin{table*}[tp]
\centering\footnotesize
\setlength{\tabcolsep}{4pt}
\renewcommand{\arraystretch}{1.0}
\caption{Retrospective evidence, part 2 of 2: attester-input forgery, insider,
frontend and liveness incidents; grouping and columns as
Table~\ref{tab:eval_per_incident}. Adshares' Q6/Q8 row is retained among the
attester-input cases.}
\label{tab:eval_per_incident_b}
\begin{tabularx}{\textwidth}{@{}>{\raggedright\arraybackslash}p{1.85cm} l X >{\raggedright\arraybackslash}p{1.55cm} c >{\raggedright\arraybackslash}p{4.6cm}@{}}
\toprule
\textbf{Bridge} & \textbf{Date} & \textbf{Failure and the question that flags it} & \textbf{Cap} & \textbf{Basis} & \textbf{Evidence} \\
\midrule
\multicolumn{6}{@{}l}{\textbf{Attester-input forgery: ingestion integrity (Q23--Q26)}}\\
THORChain Bifrost & 2021-06-29 & observer resolves inbound asset by symbol string; fake ``ETH'' credited (Q23) & Provenance & \textbf{P} & asset matched by symbol string, not source \cite{pm-thorchain-bifrost} \\
THORChain Router & 2021-07-23 & observer reads refund (\texttt{transferOut}) events as deposits (Q23) & Provenance & \textbf{P} & refund events credited as deposits \cite{pm-thorchain-router} \\
pNetwork pBTC & 2021-09-19 & enclave trusts redeem-log emitter, no canonical-source bind (Q23) & Provenance & \textbf{P} & unverified log emitter trusted as source \cite{pm-pnetwork-pbtc} \\
KelpDAO rsETH & 2026-04-18 & receive \texttt{UlnConfig} 1-of-1 DVN on-chain; poisoned RPC fails open, packet forged (Q6, Q23) & Quorum, Provenance & \correction{\textbf{C/P}} & \href{https://etherscan.io/address/0x85d456B2DfF1fd8245387C0BfB64Dfb700e98Ef3}{\texttt{0x85d4\ldots}}\,@\,\allowbreak 24908284~\allowbreak \texttt{requiredDVNCount}\,$=$\,\allowbreak 1, \correction{\texttt{balanceOf}\,$=$\,\allowbreak 116{,}723.52~rsETH} \cite{pm-kelpdao} \\
TAC TON Adapter & 2026-05-11 & sequencer credits forged Jetton-wallet deposits, releases value never locked (Q9, Q23) & Provenance, Backing & \textbf{P} & forged Jetton-wallet deposits credited \cite{pm-tac} \\
Alephium TokenBridge & 2026-05-30 & watcher accepts fabricated source events under fail-open fallback (Q23) & Provenance & \textbf{P} & fabricated source events accepted (fail-open) \cite{pm-alephium} \\
Adshares & 2026-05-15 & \texttt{wADS} minter one EOA; mint ungated by source proof (Q6, Q8) & Backing, Quorum & \textbf{C} & \href{https://etherscan.io/address/0xcfcecfe2bd2fed07a9145222e8a7ad9cf1ccd22a}{\texttt{0xcfce\ldots}}\,@\,\allowbreak 25103042~\allowbreak \texttt{isMinter}\,$=$\,\allowbreak EOA \cite{pm-adshares} \\
\addlinespace[2pt]
\multicolumn{6}{@{}l}{\textbf{Insider / rugpull: backing and mint authority (Q4, Q9)}}\\
EvoDeFi Bridge & 2022-06-07 & operator mints beyond backing (83M on 10.6M); \texttt{take()} seizes escrow (Q4, Q9) & Backing, Authority & \textbf{P} & operator over-minted; \texttt{take()} seized escrow \cite{pm-evodefi} \\
\addlinespace[2pt]
\multicolumn{6}{@{}l}{\textbf{Frontend / liveness}}\\
Celer cBridge & 2022-08-17 & hijacked config subdomain served phishing contracts; on-chain custody and attesters untouched (Q21) & \correction{None assigned} & \textbf{P} & BGP/DNS hijack served attacker addresses \cite{pm-celer} \\
Shibarium halt & 2023-08-16 & exit depended on the halted producer layer; funds stranded, not lost (Q17) & \correction{None assigned} & \textbf{P} & exit blocked when producer layer halted \cite{pm-shibarium-halt} \\
\bottomrule
\end{tabularx}
\end{table*}

\end{document}